\documentclass[12pt]{iopart}
\usepackage{iopams}
\usepackage{graphicx,bm}
\usepackage[colorlinks=true,urlcolor=blue,citecolor=blue]{hyperref}

\newcommand{\beq}{\begin{equation}}
\newcommand{\eeq}{\end{equation}}
\newcommand{\bea}{\begin{eqnarray}}
\newcommand{\eea}{\end{eqnarray}}
\newcommand{\tp}{\mathsf{T}}
\newcommand{\R}{\mathbb{R}}
\newcommand{\s}{\mathbb{S}}
\newcommand{\N}{\mathbb{N}}
\newcommand{\half}{\frac{1}{2}}
\newcommand{\bx}{\bm{x}}
\newcommand{\bX}{\bm{X}}
\newcommand{\bu}{\bm{u}}
\newcommand{\buu}{\bm{w}}
\newcommand{\be}{\bm{e}}
\newcommand{\bv}{\bm{v}}
\newcommand{\bn}{\bm{n}}
\newcommand{\mm}{\bm{m}}
\newcommand{\sgn}{\varepsilon}
\newcommand{\av}{\bX_\mathrm{av}}
\newcommand{\bw}{\bm{\omega}}
\newcommand{\V}{\mathfrak{V}}

\begin{document}

\title{Higher-order synchronization on the sphere}

\author{M~A~Lohe} 
\address{Centre for Complex Systems and Structure of Matter,
Department of Physics, The University of Adelaide, South Australia 5005, 
Australia}
\ead{Max.Lohe@adelaide.edu.au}

\begin{abstract}
We construct a system of $N$ interacting particles on the unit 
sphere $\s^{d-1}$ in $d$-dimensional space, which has $d$-body
interactions only. The equations have a gradient formulation derived from a
rotationally-invariant potential of a determinantal form summed
over all nodes, with antisymmetric coefficients. For $d=3$, for example, 
all trajectories
lie on the $2$-sphere and the potential is constructed from the triple
scalar product summed over all oriented $2$-simplices.
We investigate the cases $d=3,4,5$ in detail, and find that the system
synchronizes from generic initial values, for both
positive and negative coupling coefficients, to a static final configuration
in which the particles lie equally spaced on $\s^{d-1}$. Completely synchronized 
configurations also exist, but are unstable under the $d$-body interactions. 
We compare the relative effect of $2$-body and $d$-body
forces by adding the well-studied $2$-body interactions to the potential,
and find that higher-order interactions enhance the
synchronization of the system, specifically, synchronization 
to a final configuration consisting of equally spaced particles occurs for all
$d$-body and $2$-body coupling constants of any sign, unless
the attractive $2$-body forces 
are sufficiently strong relative to the $d$-body forces. In this case the 
system completely synchronizes as the $2$-body coupling constant 
increases through a positive critical value, with
either a continuous transition for $d=3$, or discontinuously for $d=5$.
Synchronization also occurs if the nodes have distributed natural 
frequencies of oscillation, provided that the frequencies are not 
too large in amplitude, even in the presence of repulsive 2-body interactions
which by themselves would result in asynchronous behaviour.  

\end{abstract}

\maketitle


\section{Introduction\label{intro}}

Emerging phenomena in complex systems have generally been modelled by
means of networks with interacting pairs of nodes, for which
the prevailing paradigm is the Kuramoto model \cite{K1975} 
and its many generalizations. There has been an increasing 
realization, however, that pairwise interactions and associated network
structures are not adequate to describe the group interactions 
which occur in, and possibly dominate, the behaviour of many complex systems
of interest. Higher-order interactions are known to occur in  neuroscience, 
ecology, social systems and many others \cite{Ben2016,GBM2017}. We refer to the
review \cite{Bat2020} for a detailed discussion of
higher-order interactions and their applications
with an extensive list of references, including the various 
frameworks which have been used to model higher-order systems,
as well as the recent overview \cite{Bick2021} of higher-order networks.

We describe here a hierarchy of models with higher-order
interactions which are similar to the well-known Kuramoto systems, 
and focus firstly on the properties of exclusive
$d$-body interactions, where in our formulation
$d$ is also the dimension of the system. Kuramoto models on the unit sphere
with conventional $2$-body interactions have been well-studied
and generalize the simple Kuramoto model which
corresponds to the case $d=2$. In this approach 
we associate to each of the $N$ nodes a unit vector $\bx_i$ of length $d$, 
which lies therefore on $\s^{d-1}$, and evolves
according to first-order equations in the gradient form
$\dot{\bx_i}=\nabla_i\V_d$, where the bilinear potential 
$\V_d=\V_d(\bx_1,\dots, \bx_N)$ is invariant under rotations in 
$\mathrm{SO}(d)$. Because the potential is bilinear, only $2$-body interactions
occur.

We can model higher-order interactions, however, by choosing instead 
a  multilinear determinantal form for $\V_d$ which, by antisymmetry, allows
only $d$-body interactions. For $d=3$, 
for example, for which all trajectories $\bx_i(t)$ lie on the $2$-sphere,
$\V_3$ is constructed from the triple scalar product which, being antisymmetric,
allows only $3$-body interactions, i.e.\ 
each node $i$ interacts with the other nodes only through the 
oriented $2$-simplices containing the $i^{\mathrm{th}}$ node. We find that
synchronization occurs for all values of the coupling constant, 
whether positive or negative, with the system
evolving from all initial values to the same final static configuration, 
up to an arbitrary rotation. The equations for larger values of $d$ are
of formidable complexity due to the nonlinearities
of degree $d+1$, and so we consider only $d\leqslant5$, 
although some general results may be derived.
We restrict our investigation to the simplest possible models 
with $d$-body interactions by choosing
the coupling coefficients to be signature functions, which are
antisymmetric and of magnitude either zero or one, 
but we also briefly describe the effect of oscillations 
by including distributed frequency matrices.

Besides the modelling of exclusive $d$-body forces, we can
also include pairwise interactions in any dimension $d$ by adding the usual 
bilinear terms to the potential, and so we are able to
investigate the relative effect of combined $2$-body and $d$-body forces,
in principle for any $d$, but here in detail only for $d\leqslant5$.
For $d=3$, for example, we find that the $3$-body forces 
enhance synchronization, and so if 
the coupling constant for the $2$-body interactions is negative,
resulting in repulsive $2$-body forces that
would otherwise prevent the system from synchronizing, 
the combined system nevertheless synchronizes due to the 3-body interactions.
This is consistent with previous observations 
that ``higher-order interactions can stabilize strongly 
synchronized states even when the pairwise coupling is repulsive" 
\cite{GBM2017,Sk2020}.

For a general description of synchronization with respect to higher-order
interactions we refer to \cite{Bat2020} (Section 6) also
\cite{Bick2021} (Section 4) and \cite{Port2020}, 
and note that extensions of the Kuramoto model to
higher-order networks have been extensively investigated, but generally 
with trajectories restricted to the unit circle $\s^1$ 
\cite{Tan2011,Sk2019,Xu2020,Sk2021,G2021,Millan2020}. 
For a description and properties of simplicial complexes, and the relation to 
algebraic topology we refer to \cite{Bat2020,G2021,Vil2021},
however our focus here is on the properties of  specific dynamical systems 
defined on the unit sphere which allow higher-order interactions, and
their synchronization behaviour. The effect
of $2$-body forces in combination with $3$- and $4$-body forces 
has previously been considered, but only in the context of phase 
oscillators on $\s^1$ \cite{Sk2020,Tan2011}.
Properties such as multistability, which refers to the
coexistence of multiple steady states, has been observed
in these and other models \cite{Tan2011,Sk2019,Vil2021}, and
it is generally thought 
``that higher-order interactions favor multistability" \cite{Bat2020}, but
it remains an open question as to whether such properties
are model-specific, or are general consequences of higher-order interactions.
Multistability, for example, does not appear in the models that we consider,
although to any stable steady state solution there corresponds, by 
rotational covariance, a family of rotated steady states.
A similar observation can be made with respect to ``explosive synchronization" 
which has been observed in a higher-order Kuramoto model \cite{Millan2020},
although similar properties are well-known to occur in models
with only $2$-body interactions \cite{FPK2016,Kumar2021,DS2019}. 
This behaviour is also absent in the higher-order models that we 
consider, similarly
for the property of  ``abrupt transitions" as appears in various models 
\cite{Bat2020,Sk2020,Sk2019}. We do find, however, that there is a 
discontinuous transition for $d=5$ at which the $2$-body forces suddenly become 
sufficiently strong to overcome the 
5-body forces, as discussed in Section \ref{s6}.
One difference in our approach, which possibly eliminates
discontinuous behaviours, is that our equations take a gradient
form which can be used to  deduce the stability of steady states 
of the model, as explained in \cite{Vil2021}.


\subsection{Outline and summary of results}

We begin in Section \ref{s2} by discussing models of synchronization on 
the unit sphere which form a 
natural generalization of the widely-studied Kuramoto model. 
We formulate the equations
as a gradient system, choosing firstly the familiar bilinear
potential in Section \ref{ss21} which leads to pairwise interactions, 
followed in Section \ref{ss22} by a multilinear
potential with antisymmetric couplings, which leads
to $d$-body interactions on $\s^{d-1}$. Whereas the $2$-body
equations have cubic nonlinearities with $N-1$ summations at each node,
corresponding to $2$-body interactions with the other $N-1$ nodes,
the $d$-body systems have nonlinearities of order $d+1$, with 
$(N-1)!/(N-d)!$ summations at each node, which interacts
with the other nodes by means of a connected $(d-1)$-simplex. 
It is straightforward to combine the $2$-body and $d$-body systems,
and so we investigate the relative effect of the two couplings with
corresponding strengths $\kappa_2,\kappa_d$.
We discuss steady state solutions and their stability in Section
\ref{ss24} for both the $d$-body system and the combined $2$-body and
$d$-body systems. 
The stable steady state configurations for $d$-body systems consist
of equally spaced nodes on $\s^{d-1}$, rather than completely synchronized
configurations as occur for $2$-body couplings. For combined systems,
the $d$-body steady states prevail, unless
the ratio of coupling constants  $\kappa_2/\kappa_d$ is sufficiently large 
to favour the $2$-body forces.

In Section \ref{s4} we focus on the case $d=3$, this being
the simplest of the higher-order models, 
although $d=2$ is also of interest but is considered later 
in Section \ref{s3} because it has only $2$-body interactions. 
We show that the $d=3$ system does not
completely synchronize, as occurs for the $2$-body models on the sphere,
rather for generic initial values the nodes arrange themselves asymptotically 
in a ring formation.
Exactly why ring synchronization, rather than complete synchronization,
occurs is discussed for the special case $N=3$ in Section \ref{ss43}, 
for which we can solve the nonlinear
equations exactly. We consider combined $2$- and 3-body interactions 
in Section \ref{sec4}, obtaining an exact expression for the
steady states which again form an equally spaced ring of nodes on $\s^2$.
These states exist and are stable, unless the ratio of coupling constants
$\kappa_2/\kappa_3$ becomes sufficiently large, at which point the system 
transitions continuously to a completely synchronized formation.
We can state precisely the value of $\kappa_2/\kappa_3$ at which this occurs,
see Section \ref{ss42}.

Properties of the general system depend on whether $d$ is even or odd,
and so we have selected two even cases $d=2,4$ and two odd cases $d=3,5$
for detailed investigation. For even $d$, the equally spaced synchronized nodes
do not form a closed sequence, as occurs for odd $d$, as we see
in Section \ref{s5} for $d=4$.  The 
asymptotic configurations for the combined $2$- and $4$-body systems, discussed
in Section \ref{ss52}, are of a form that is not easily 
analyzed, but
general arguments show that again synchronization is controlled by the 
4-body interactions, unless the ratio $\kappa_2/\kappa_4$
becomes sufficiently large, then the attractive forces due to the 
$2$-body interactions force the system to completely synchronize.
For the 5-body system, which is similar to the case $d=3$, we obtain
exact steady state expressions which extend to the combined system
with $2$-body interactions, see Section \ref{s6}. In this case there is 
a discontinuous transition to a completely synchronized system at a critical
ratio $\kappa_2/\kappa_5$.

Finally we consider the $d=2$ case in Section \ref{s3}, which we have left
to last because only $2$-body interactions are involved, however the system
has properties which differ from those of the standard Kuramoto model,
due to the antisymmetric coupling constants.
There is merit in investigating this as the simplest of the
$\s^{d-1}$ models and also as a guide to the behaviour of the systems for $d>2$.
Again, in the combined system the antisymmetric couplings enhance the 
synchronization of the Kuramoto model.


\section{Synchronization models on the unit sphere\label{s2}}

Synchronization models in which the $N$ trajectories of
the complex system are confined to the unit sphere have been 
extensively investigated, particularly for the special case of the
Kuramoto model. In general, a unit $d$-vector $\bx_i$ is 
associated with the $i^{\mathrm{th}}$ node of the network
and evolves according to the nonlinear equations obtained
by minimizing a  potential  $\V_d(\bx_1,\dots, \bx_N)$, where
the constraint $\bx_i\centerdot\bx_i=1$ for every
$i=1,\dots N$ is enforced by means of Lagrange multipliers.  

The first-order evolution equations have the gradient form
$\dot{\bx_i}=\nabla_i\V_d$, to which one can add further terms 
such as $\Omega_i\bx_i$, where $\Omega_i$ is a 
$d\times d$ antisymmetric frequency matrix, in which case each node 
has one or more natural modes 
of oscillation as determined by the eigenvalues of $\Omega_i$.
Properties of the model depend on the choice of $\V_d$,
which is invariant under transformations of the
rotation group $\mathrm{SO}(d)$. There are two possible such choices for $\V_d$,
either the bilinear inner product, or the triple scalar product for $d=3$
and its generalization to any $d$. This leads to two types of models,
those with pairwise  interactions, and those with $d$-body interactions. 
We can also additively combine these two possibilities and hence investigate 
the relative effect of $d$-body and $2$-body forces within the system.


\subsection{Pairwise interactions on the unit sphere\label{ss21}}

The choice for $\V_d$ which has been widely investigated is
\beq
\label{e1}
\V_d=\sum_{i,j=1}^N a_{ij}\,\bx_i\centerdot\bx_j,
\eeq
where the coefficients $a_{ij}$ are symmetric, and in this case the
evolution equations are
$\dot{\bx_i}=-\lambda_i\bx_i+\sum_j a_{ij}\bx_j$
where $\lambda_i$ are Lagrange multipliers,
supplemented by the constraint $\bx_i\centerdot\bx_i=1$. Hence 
$\lambda_i=\sum_j a_{ij} \bx_i\centerdot\bx_j$ and so we obtain the
system of $N$ equations
\beq
\label{e2}
\dot{\bx_i}
=
\Omega_i\bx_i
+\frac{\kappa_2}{N}
\sum_{j=1}^N a_{ij}\left[\bx_j-(\bx_i\centerdot\bx_j)\;\bx_i\right],
\eeq
where we have included $d\times d$ antisymmetric frequency matrices $\Omega_i$,
as well as a normalized coupling coefficient $\kappa_2$ which measures
the relative strength of the natural oscillations compared
to the nonlinear interactions. Since $\bx_i\centerdot\dot{\bx_i}=0$, the unit
length of $\bx_i$ is maintained as the system evolves.
The Kuramoto model corresponds to the 
case $d=2$ with the parametrization $\bx_i=(\cos\theta_i,\sin\theta_i)$
and $\Omega_i=\pmatrix{0 &-\omega_i \cr\omega_i &0}$,
for which (\ref{e2}) reduces to
$
\dot{\theta_i}
=
\omega_i+\frac{\kappa_2}{N}\sum_{j=1}^Na_{ij}\sin(\theta_j-\theta_i)$.
In this case the  potential  (\ref{e1}) reduces to 
$\V_2=\sum_{i,j}a_{ij}\cos(\theta_j-\theta_i)$ which is well-known
as a Lyapunov function for the Kuramoto model \cite{HW1993,MS2005}.
It is convenient to write (\ref{e2}) in the form
\beq
\label{f4}
\dot{\bx_i}
=
\Omega_i\bx_i+\bX_i-\bx_i\;(\bx_i\centerdot\bX_i),
\eeq
where $\bX_i=\kappa_2 \sum_{j=1}^N a_{ij}\bx_j/N$, a form which is maintained
also for the $d$-body system. For global
coupling, $a_{ij}=1$ for all $i,j$, every node connects to the average
position $\av$ defined by
\beq
\label{f9}
\av=\frac{1}{N}\sum_{j=1}^N\bx_j,
\eeq
then we can write (\ref{e2}) as:
\beq
\label{f6}
\dot{\bx_i}
=
\Omega_i\bx_i+\kappa_2\av-\kappa_2\,\bx_i\,(\bx_i\centerdot\av).
\eeq

The system (\ref{e2}) has been intensively investigated for
general $d$, usually with global coupling
and for the homogeneous case $\Omega_i=0$,
whether as a model of synchronization
\cite{ML2009,CH2014,CCH2014,ZZ2019,Mark2020,JC2020,ML2020}, 
or opinion formation and
consensus studies on the unit sphere 
\cite{OS2004,LS2014,Cap2015,LS2016,Zhang2018,MJG2018,CJ2020},
or for the modelling of swarming behaviour 
\cite{OS2006,MPG2020,Chandra2019}.
Many synchronization properties have been established for any $d$
\cite{CH2014,CCH2014,ZWZ2019,HKLN2020,HK2021}, in particular
for the case of identical frequencies $\Omega_i=\Omega$, 
it is known that for $\kappa_2>0$ the order parameter $r=\|\av\|$ evolves
exponentially quickly  to the value $r_{\infty}=1$, 
in which case all nodes
are co-located to form a completely synchronized state, or for $\kappa_2<0$
to a state with $r_{\infty}=0$. Indeed, it is clear from (\ref{f6}) that 
for $\Omega_i=0$ either $\bx_i=\bu=\av$ for some fixed unit vector $\bu$,  or 
$\av=0$, each provide a steady state solution. 
For non-identical matrices $\Omega_i$
with a sufficiently large positive coupling $\kappa_2$, the system
undergoes ``practical synchronization'' \cite{CCH2014} as 
$\kappa_2\to\infty$, meaning that
the configuration becomes asymptotically close to a completely
synchronized state \cite{HKLN2020}. If $\kappa_2<0$, with distributed
frequency matrices $\Omega_i$, the system remains asynchronous, a well-known
property of the Kuramoto model.
For specific values of $d$ more precise properties can be proved, 
for example for the
Kuramoto model phase-locked synchronization occurs for any
$\kappa_2>\kappa_c$ for some critical coupling $\kappa_c>0$
\cite{CHJK2012,Dong2013,HKP2015}. 
The $d=4$ case, for which trajectories lie on $\s^3$, equivalently
on the $\mathrm{SU}_2$ group manifold, is similar to $d=2$, as numerical 
studies show \cite{ML2009}.


\subsection{Higher-order interactions on the 
unit sphere\label{ss22}}

The unit sphere models which follow from the  potential  (\ref{e1})
have only pairwise interactions, and so we now consider models with 
higher-order interactions. 
Let $\N_N=\{1,2,\dots N\}$, with $N\geqslant d$, then 
$(\bx_{i_1},\bx_{i_2},\dots, \bx_{i_d})$
is a $d\times d$ matrix, where $\bx_i$ is a unit 
column vector of length $d$, and $i_1,i_2,\dots i_d\in\N_N$. 
Define the  potential 
\beq
\label{e4}
\V_d=\sum_{i_1,i_2,\dots i_d=1}^N a_{[i_1,i_2,\dots i_d]}\,
\det(\bx_{i_1},\bx_{i_2},\dots, \bx_{i_d}),
\eeq
where the summation is over all  $i_1,i_2,\dots i_d\in\N_N$ and
the coefficients $a_{[i_1,i_2,\dots i_d]}$ are antisymmetrized over 
these indices, corresponding to the antisymmetry of the determinant. For 
$d=3$, for example, we antisymmetrize any set of coefficients
$a_{ijk}$ according to
\[
a_{[i,j,k]}=\frac{1}{3!}(a_{ijk}+a_{kij}+a_{jki}-a_{jik}-a_{ikj}-a_{kji}).
\]
$\V_d$ as defined by (\ref{e4}) is rotationally invariant, i.e.\
if $\bx_i\to R\bx_i$ where $R\in\mathrm{SO}(d)$, then
$\det(\bx_{i_1},\bx_{i_2},\dots, \bx_{i_d})$ remains unchanged
because $\det R=1$. If, however, $R\in\mathrm{O}(d)$ with $\det R=-1$,
then $\V_d$ changes sign, for example if $d$ is odd, then $\V_d$ changes sign
under parity inversion $\bx_i\to-\bx_i$. As a more general example, 
for any $d$, we 
can choose $R$ to be the identity matrix but with the sign of any 
one diagonal element reversed, then $R\in\mathrm{O}(d)$ with $\det R=-1$, 
and the sign of $\V_d$ is reversed.

The  potential (\ref{e4}) leads to a hierarchy of synchronization models
on the
unit sphere $\s^{d-1}$ with $d$-body couplings, since interactions 
take place between nodes only as constituents of oriented $(d-1)$-simplices. 
The order of the interaction 
is evidently tied to the dimension $d$ of the vector $\bx_i$, 
and hence to the dimension of the unit sphere $\s^{d-1}$.
For $d=2$ we write
$\bx_i=(\cos\theta_i,\sin\theta_i)$ and so
$\V_2=\sum_{i,j=1}^Na_{[ij]}\sin(\theta_j-\theta_i)$, which we consider 
in Section \ref{s3}, and for $d=3$ we obtain 
\beq
\label{f5}
\V_3
=
\sum_{i,j,k=1}^Na_{[ijk]}\det(\bx_i,\bx_j,\bx_k)
=
\sum_{i,j,k=1}^Na_{[ijk]}\;\bx_i\centerdot\bx_j\times\bx_k,
\eeq
to be analyzed in Section \ref{s4}.

The coefficients $a_{[i_1,i_2,\dots i_d]}$ in (\ref{e4}) are antisymmetric 
but are otherwise unrestricted and so, in order to investigate the 
simplest possible case, and
by analogy  with the symmetric coupling $a_{ij}=1$ which we imposed for the 
pairwise interactions in the system (\ref{e2}), we choose
the coefficients $a_{[i_1,i_2,\dots i_d]}$ to be the 
signature, denoted  $\sgn_{i_1i_2\dots i_d}$, of the permutation
$P=\{i_1,i_2,\dots i_d\}$. The signature is defined as the parity 
$(-1)^n$ of $P$, where $n$ is the 
number of transpositions of pairs of elements that must be composed to 
construct the permutation. This definition extends to indices
which are not permutations by setting the value to be zero.
We have, for example, $\sgn_{ij}=1$ if $j>i$, $\sgn_{ij}=-1$ if $j<i$, 
and $\sgn_{ij}=0$ if $j=i$ for any $i,j\in\N_N$.
Similarly, $\sgn_{ijk}=1$ if $i<j<k$, or for any cyclic permutation
of $i,j,k$, $\sgn_{ijk}=-1$ for any anticyclic permutation, and zero 
if any two indices are equal. In these higher-order models, therefore,
each node interacts with all the other nodes in a connected $(d-1)$-simplex,
with equal strength in magnitude, by means of the
coupling coefficients $\sgn_{i_1i_2\dots i_d}$, with a sign 
depending on the orientation. There are $N!/(N-d)!$ nonzero coefficients
$\sgn_{i_1i_2\dots i_d}$, and hence there are
$N!/(N-d)!$ independent $d$-body couplings between the $N$ nodes,
which take place through the $(d-1)$-simplices.


\subsection{Nonlinear $d$-body equations on the sphere}

In order to calculate the gradient of $\V_d$ as defined
in (\ref{e4}), we write the determinant in terms of the
Levi-Civita symbol $\varepsilon^{a_1a_2 \dots a_d}$ and the components
$x_i^a$ of $\bx_i$, as follows:
\beq
\label{e5}
\V_d
=
\sum_{i_1,i_2,\dots i_d=1}^N 
\sum_{a_1,a_2, \dots a_d=1}^d 
\sgn_{i_1i_2\dots i_d}\;
\varepsilon^{a_1a_2 \dots a_d}\, x_{i_1}^{a_1}x_{i_2}^{a_2}\dots x_{i_d}^{a_d}.
\eeq
For $d=3$, for example, we have
$\det(\bx_i,\bx_j,\bx_k)=\bx_i\centerdot\bx_j\times\bx_k
=
\sum_{a,b,c=1}^3 \varepsilon^{abc}x_i^ax_j^bx_k^c$. For each set of indices
$i_2\dots i_{d}$ define the
vector $\bv_{i_2\dots i_{d}}$ of length $d$ with components
\beq
\label{f7}
(\bv_{i_2\dots i_{d}})^a
=
\sum_{a_2\dots a_d=1}^d 
\varepsilon^{a\,a_2 \dots a_d}\, x_{i_2}^{a_2}\dots x_{i_d}^{a_d},
\eeq
then $\bv_{i_2\dots i_{d}}$ behaves as a vector under
rotations, i.e.\ if  $\bx_i\to R\bx_i$ where $R\in\mathrm{SO}(d)$, then 
$\bv_{i_2\dots i_{d}}\to R\bv_{i_2\dots i_{d}}$.
If $R\in\mathrm{O}(d)$ with $\det R=-1$, however, then
$\bv_{i_2\dots i_{d}}\to -R\bv_{i_2\dots i_{d}}$.
 For $d=3$ we have $\bv_{ij}=\bx_i\times\bx_j$. The identity
$\bu\centerdot\bv_{i_2\dots i_{d}}= \det(\bu,\bx_{i_2},\dots, \bx_{i_d})$,
for any fixed vector $\bu$,
provides a convenient method for calculating $\bv_{i_2\dots i_{d}}$.
In particular, we have
$\bx_i\centerdot\bv_{i_2\dots i_{d}}= \det(\bx_i,\bx_{i_2},\dots, \bx_{i_d})$.

Vectors such as $\bv_{i_2\dots i_{d}}$ can be viewed as the dual of 
elements belonging to an exterior algebra on which is defined 
a wedge or exterior product. 
The dual vector, generally referred to as the Hodge dual, has components formed
using the Levi-Civita symbol as shown in (\ref{f7}). In three dimensions,
for example, the vector product of two vectors can be viewed as the dual
of the wedge product of these vectors. For our purposes, particularly
for numerical evaluation or with a computer algebra system, 
it is sufficient to determine properties and explicit expressions for  
$\bv_{i_2\dots i_{d}}$ either directly from (\ref{f7}) or by means
of $\bu\centerdot\bv_{i_2\dots i_{d}}= \det(\bu,\bx_{i_2},\dots, \bx_{i_d})$,
which holds for any fixed $d$-vector $\bu$.
For a detailed description of exterior algebras, properties of 
the antisymmetrization operator, and the definition
of the Hodge dual, we refer to \cite{Sz}, Chapter 8.

The gradient of the potential $\V_d$ defined by (\ref{e5}) is given
for unconstrained vectors $\bx_i$ by:
\[
\nabla_i\V_d
=
d \sum_{i_2,\dots i_d=1}^N \sgn_{i, i_2\dots i_d}\;
\bv_{i_2\dots i_{d}},
\]
and so after including Lagrange multipliers in order to enforce 
the constraints, the equations of motion are:
\beq
\label{e6}
\dot{\bx_i}
=
\frac{\kappa_d}{N^{d-1}} \sum_{i_2,\dots i_d=1}^N \sgn_{i, i_2\dots i_d}
\left[\bv_{i_2\dots i_{d}}-\bx_i\,(\bx_{i}\centerdot\bv_{i_2\dots i_{d}})
\right],
\eeq
where we have included a normalized coupling constant $\kappa_d$.
For fixed $i$ there are $(N-1)!/(N-d)!$ nonzero terms under the summation,
which for large $N$ approaches $N^{d-1}$, and so we have normalized
$\kappa_d$ by this factor. We can add oscillator
terms $\Omega_i\bx_i$ to the right-hand side, as for the system
(\ref{e2}), but if the matrices $\Omega_i$ are identical,
$\Omega_i=\Omega$, then we can replace $\bx_i\to\rme^{\Omega t}\bx_i$
in the usual way and so because $\rme^{\Omega t}$ is a rotation matrix, 
we can in effect set $\Omega=0$. In this case the coefficient
$\kappa_d/N^{d-1}$ in (\ref{e6}) can be set to $\pm1$ by rescaling the 
time variable. For every solution of (\ref{e6}) with $\kappa_d>0$
there corresponds another solution with $\kappa_d<0$, obtained by
transforming $\bx_i\to R\bx_i$, where $R\in\mathrm{O}(d)$ with $\det R=-1$.
If $d$ is odd, for example, parity inversion $\bx_i\to-\bx_i$ in effect
changes the sign of $\kappa_d$. The properties of the system (\ref{e6})
are therefore independent of the sign of $\kappa_d$, in contrast to the
Kuramoto model, or more generally the $2$-body system  (\ref{e2}).
The form  of (\ref{e6}) is the same as shown in (\ref{f4}), except
that now each vector $\bx_i$ couples to
$\bX_i= \kappa_d\sum_{i_2,\dots i_d=1}^N 
\sgn_{i, i_2\dots i_d}\bv_{i_2\dots i_{d}}/N^{d-1}$, which
necessarily depends on $i$, and so the coupling takes place through
all possible $(d-1)$-simplices containing the $i^{\mathrm{th}}$ node.

We wish to determine the behaviour of the system (\ref{e6}) for
generic initial values $\bx_i(0)=\bx_i^0$, where ``generic" means we
exclude unstable fixed points, and look for
synchronized final configurations.  A useful measure of synchronization, 
whether for $2$-body or $d$-body interactions,
is by means of the average position $\av$ defined in (\ref{f9}),
from which we calculate the rotationally invariant order parameter 
$r=\|\av\|$. Synchronization
occurs when $r$ achieves a constant asymptotic value $r_{\infty}$,
where $0\leqslant r_{\infty}\leqslant1$. If $r_{\infty}=1$ all 
particles are co-located at a common point, usually referred to as a 
completely synchronized configuration, 
and if $r_{\infty}=0$ then $\av=0$ as $t\to\infty$, hence
the particles are distributed over the sphere with an average position of zero,
sometimes referred to as a balanced configuration.
Both cases occur for the 2-body system (\ref{e2}) with identical 
frequencies, depending on the sign of the coupling \cite{CH2014}.


\subsection{Steady state solutions\label{ss24}}

For the  values of $d$ under consideration we find numerically
that the system (\ref{e6})
always attains a final static configuration for all generic initial values, 
and so we wish to determine all stable steady state solutions.
This appears to be a formidable task,
considering that the nonlinearities on the right-hand side of
(\ref{e6}) are of order $d+1$, however we  show that the static
equations can be solved by a simpler reduced system. 
Due to the rotational covariance of (\ref{e6}), any static configuration 
can be rotated to an arbitrary orientation, and so for numerical comparison 
of final configurations, which always depend on the initial values, 
we calculate rotational invariants such as  $\bx_i\centerdot\bx_j$.

We observe firstly that (\ref{e6})
has the fixed point solution $\bx_i=\pm\bu$, where $\bu$ is any constant 
unit vector, with a sign that can depend on $i$, because by antisymmetry 
$\bv_{i_2\dots i_{d}}$ is zero if the vectors $\bx_i$ are either
parallel or antiparallel.
There is the possibility therefore that the system
completely synchronizes, as occurs for the $2$-body system
(\ref{e2}), however numerically we find that such
configurations are unstable. Instead, we  look for static 
solutions satisfying
\beq
\label{e12}
\sum_{i_2,\dots i_d=1}^N \sgn_{i, i_2\dots i_d}
\bv_{i_2\dots i_{d}}
=\lambda\, \bx_i,
\eeq
where $\lambda$ is independent of $i$, and therefore takes the value
\[
\lambda
=
\sum_{i_2,\dots i_d=1}^N \sgn_{i, i_2\dots i_d}
\bx_i\centerdot\bv_{i_2\dots i_{d}}
=
\sum_{i_2,\dots i_d=1}^N \sgn_{i, i_2\dots i_d}
\det(\bx_i,\bx_{i_2},\dots, \bx_{i_d}).
\]
It follows from (\ref{e12}) that the right-hand side of 
(\ref{e6}) is zero, indeed  this is true
even if $\lambda$ depends on $i$, however we find numerically
that static solutions satisfy the special form (\ref{e12}) with 
$\lambda$ independent of $i$, and we show by direct calculation
that this holds exactly for $d=3$ see (\ref{e17}),  for $d=4$ see
(\ref{e35}) and for $d=5$ see (\ref{f50}).

We are also interested in combining $2$-body and $d$-body systems in order 
to evaluate the relative effect of the corresponding couplings, hence
we combine (\ref{f6}) and (\ref{e6}) to obtain
\beq
\label{g13}
\fl
\dot{\bx}_i
=
\kappa_2\av-\kappa_2\,\bx_i(\bx_i\centerdot\av)
+
\frac{\kappa_d}{N^{d-1}} \sum_{i_2,\dots i_d=1}^N \sgn_{i, i_2\dots i_d}
\left[\bv_{i_2\dots i_{d}}-\bx_i\,(\bx_{i}\centerdot\bv_{i_2\dots i_{d}})
\right],
\eeq
where $\kappa_2$ denotes the strength of the $2$-body forces.
We can satisfy the static equations in this case by solving
\beq
\label{g14}
\sum_{i_2,\dots i_d=1}^N \sgn_{i, i_2\dots i_d}\bv_{i_2\dots i_{d}}
=
\lambda_1\bx_i-\lambda_2\av,
\eeq
for constants $\lambda_1,\lambda_2$ independent of $i$, since then
we have
\[
\sum_{i_2,\dots i_d=1}^N \sgn_{i, i_2\dots i_d}
\bx_{i}\centerdot\bv_{i_2\dots i_{d}}
=
\lambda_1-\lambda_2\,(\bx_{i}\centerdot\av),
\]
and hence the right-hand side of  (\ref{g13}) is zero,
provided that
\beq
\label{f15}
\lambda_2=
\frac{\kappa_2 N^{d-1}}{\kappa_d}.
\eeq
More directly, define
\beq
\buu_i=
\frac{\kappa_d}{N^{d-1}}\sum_{i_2,\dots i_d=1}^N \sgn_{i, i_2\dots i_d}\bv_{i_2\dots i_{d}}
-
\frac{\lambda_1\kappa_d}{N^{d-1}}\bx_i+\kappa_2\av,
\eeq
then we can rewrite (\ref{g13}) equivalently as
$\dot{\bx}_i
=
\buu_i-\bx_i\,(\bx_{i}\centerdot \buu_i)$,
and so we see immediately that $\dot{\bx}_i=0$ for $\buu_i=0$.
We note here that the signs of $\lambda_1,\lambda_2$ in (\ref{g14})
are each reversed under the
transformation $\bx_i\to R\bx_i$, where $R\in\mathrm{O}(d)$ with $\det R=-1$,
which is equivalent to reversing the sign of $\kappa_d$, and so the signs of 
$\lambda_1,\lambda_2$ are correlated with the sign of $\kappa_d$.

It is not at all evident that stable static solutions should
satisfy (\ref{g14}) for all $i$, but
numerically we find this to be true in all cases, 
and prove for $d=3,5$ that it holds 
exactly. For static solutions that depend on one or more unknown
parameters, we can in
principle determine $\lambda_2$ as a function of these parameters, then
(\ref{f15}) expresses these parameters explicitly  in terms of the ratio
$\kappa_2/\kappa_d$. Let us proceed now to the case $d=3$, for which
we can find exact steady state solutions and so verify (\ref{g14}).


\section{3-body interactions on the $2$-sphere\label{s4}}

For $d=3$ the $N$ equations (\ref{e6}) for the unit 3-vectors 
$\bx_i\in\s^2$ are:
\beq
\label{e13}
\dot{\bx_i}
=
\bw_i\times\bx_i 
+
\frac{\kappa_3}{N^2}\sum_{j,k=1}^N\sgn_{ijk}\,
\left[\bx_j\times\bx_k-\bx_i\,(\bx_i\centerdot\bx_j\times\bx_k)\right],
\eeq
where we have included frequency 3-vectors 
$\bw_i=(\omega_i^1,\omega_i^2,\omega_i^3)$.

\begin{figure}[!ht]
\begin{center}
\includegraphics[width=0.49\columnwidth,trim={40 80 120 60},clip]{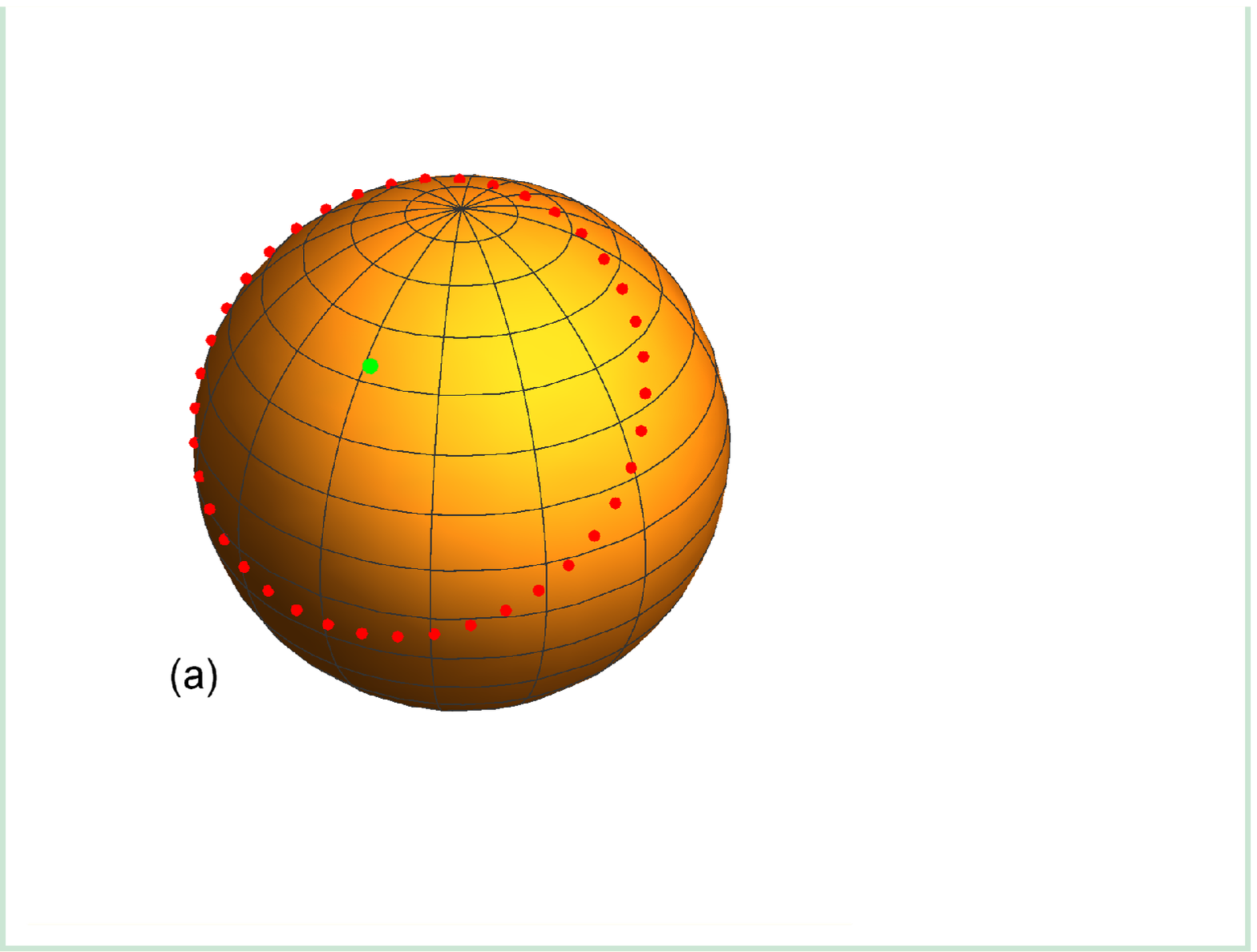}
\includegraphics[width=0.49\columnwidth,trim={40 80 120 60},clip]{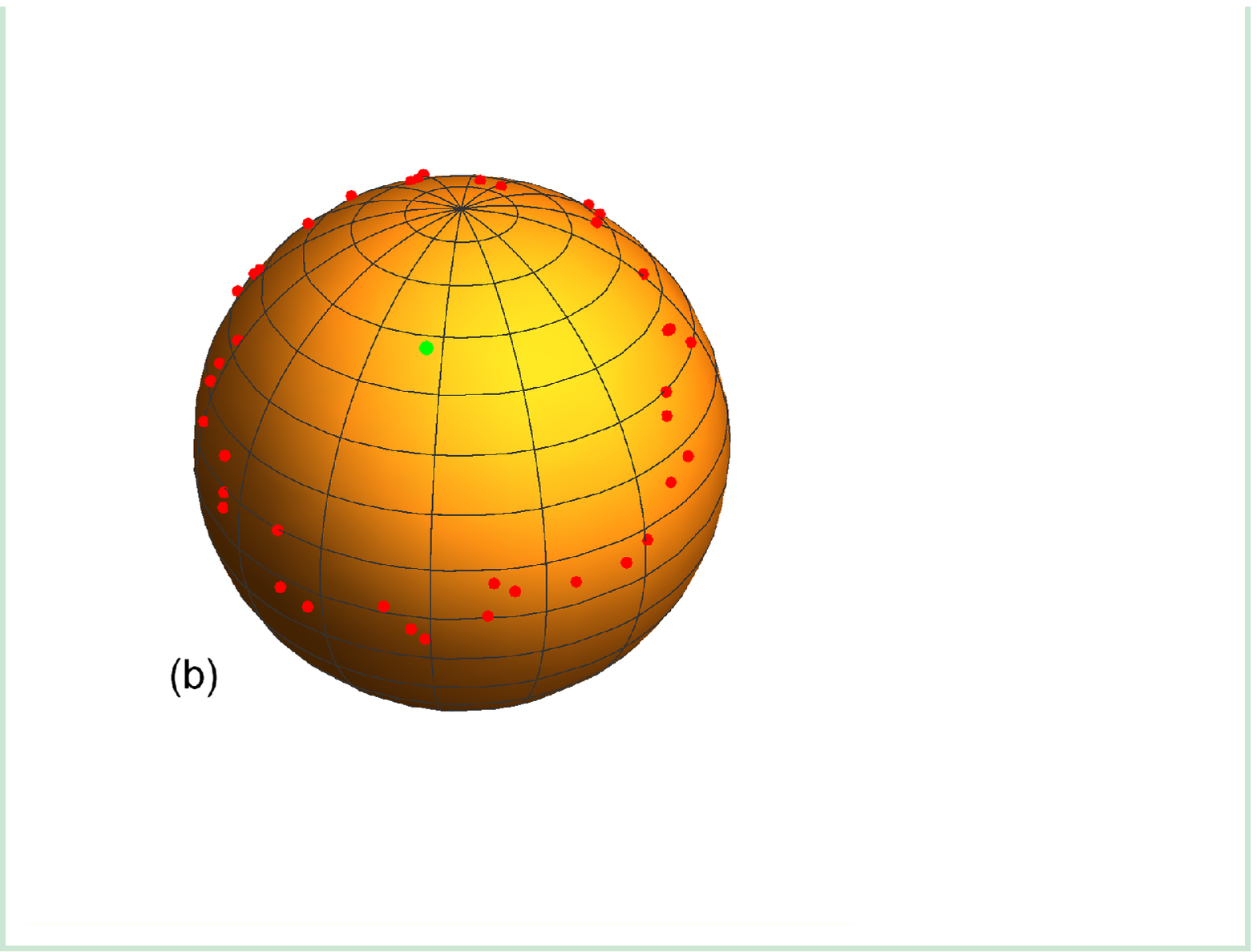}
\end{center}
\caption{
(a) An example of ring synchronization for the system (\ref{e13}) 
with $\bw_i=0$ for $N=40$, showing the unit normal (green) to the 
plane containing the ring of nodes;
(b) a synchronized configuration with distributed frequency 
vectors $\bw_i$ with $\|\bw_i\|\approx1$ for $N=40, \kappa_3=20$, 
showing approximate ring synchronization.
}
\label{fig1}
\end{figure}

For the homogeneous case $\bw_i=0$,
the right-hand side of (\ref{e13}) has a gradient form derived from 
the  potential 
$\V_3=\sum_{i,j,k=1}^N\sgn_{ijk}\;\bx_i\centerdot\bx_j\times\bx_k$, and
the coupling constant $\kappa_3/N^2$ can be scaled to unity, with a
sign that can be reversed by means of the parity transformation $\bx_i\to-\bx_i$ 
for all $i$.  The system has the fixed point $\bx_i=\bu$, 
where $\bu$ is any constant unit vector but, numerically, we find that 
such completely synchronized configurations are unstable.
Instead, the system synchronizes from 
generic initial values to a static configuration,
with an orientation that depends on the initial values, 
of the form shown in Fig.\ \ref{fig1}(a). We refer to this
as ring synchronization, since all nodes lie equally spaced on a circle 
arising from the
intersection of a plane with the unit sphere. Fig.\
\ref{fig1}(a) shows the configuration for $N=40$ nodes (red), 
and the unit normal $\bn$ (green), where $\bn=\av/\|\av\|$ with 
$\av$ defined in (\ref{f9}). The asymptotic configuration in all cases
satisfies $\bn\centerdot\bx_i=\frac{1}{\sqrt{3}}$ for all
$i$, showing that all nodes lie in the plane 
$\bn\centerdot\bx=\frac{1}{\sqrt{3}}$, where $\bx=(x,y,z)$, from which
it follows that $\|\av\|=r_{\infty}=\frac{1}{\sqrt{3}}$. Unlike the cases
 $d=2,4$, see (\ref{e9},\ref{g37}) respectively,
$r_{\infty}$ is independent of $N$.

We can write down an explicit expression for ring-synchronized
configurations such as in Fig.\ \ref{fig1}(a), up to an 
arbitrary  $\mathrm{SO}(3)$ rotation, and show that these are exact fixed 
point solutions of (\ref{e13}). The stability of these configurations
is a numerical observation, however, 
although we prove this to be the case for 
$N=3$ in Section \ref{ss43} by solving exactly for the
rotational invariants.

\subsection{Steady state solutions on $\s^2$\label{ss41}}

For any static synchronized configuration as shown in Fig.\ \ref{fig1}(a) we can 
point the normal $\bn$ along the $Z$-axis, i.e.\ we can set 
$\bn=(0,0,1)$ by means of an $\mathrm{SO}(3)$ rotation. 
Such rotations can be performed algebraically or numerically 
as explained in \ref{C}. Then, since 
all nodes lie in the plane $z=\frac{1}{\sqrt{3}}$, 
the steady state solution is given by:
\beq
\label{e16}
\bx_i=
\frac{1}{\sqrt{3}}\left(\sqrt{2}\cos\frac{2i\pi}{N},
\sqrt{2}\sin\frac{2i\pi}{N},1\right),
\eeq
for $i=1,\dots N$,
where we have also performed an
$\mathrm{SO}(2)$ rotation about the $Z$-axis so that 
$\bx_N=\frac{1}{\sqrt{3}}\left(\sqrt{2},0,1\right)$ is aligned along the
$X$-axis. The sign of $\bx_i$ varies according to the sign of $\kappa_3$, i.e.\   
either (\ref{e16}) or its negative is a  stable fixed point.

From (\ref{e16}) we can compute the following rotational invariants,
where we denote 
$x_{ij}=\bx_i\centerdot\bx_{j}$ and $x_{ijk}=\bx_i\centerdot\bx_j\times\bx_k$: 
\bea
\label{f12}
x_{ij}&=&\frac{1}{3}\left[1+2\cos\frac{2(i-j)\pi}{N}\right],
\\
\label{f13}
x_{ijk}
&=&
\frac{8}{3\sqrt{3}}\sin\frac{(i-j)\pi}{N}
\sin\frac{(k-i)\pi}{N}\sin\frac{(j-k)\pi}{N}.
\eea
These invariants are related by the identity
$x_{ijk}^2=1-x_{ij}^2-x_{ik}^2-x_{jk}^2+2x_{ij}x_{ik}x_{jk}$ as we discuss
for $N=3$ in Section \ref{ss43}. 
Since these expressions are independent of the orientation of the
system, they provide a convenient means for the numerical verification
of all exact expressions,
for any initial values, and hence for all final configurations.

We see from (\ref{f12}) that $x_{ij}$ depends only on the difference
$|i-j|$, one consequence of which is that adjoining nodes are equally
spaced, since $\|\bx_i-\bx_{i+1}\|$ is independent of $i$. In addition,
this distance is equal to $\|\bx_1-\bx_{N}\|$, showing that
the nodes form a closed loop, as follows from the symmetry
$\bx_i\centerdot\bx_{N}=\bx_{N-i}\centerdot\bx_{N}$ for $i=1,2 \dots N-1$.
Although this property is evident from Fig.\ \ref{fig1}(a), it  
does not hold for even values of $d$, see Section \ref{s5} for $d=4$. 
It is convenient to extend the formula (\ref{e16}) to the value
$i=0$, then we have $\bx_0=\bx_N$ which again shows that the nodes form
a closed sequence.
Configurations with equally spaced nodes
are sometimes referred to as splay states, although for $d>3$
the asymptotic configurations are not restricted to a plane as occurs here;
for $d=4$, for example, we obtain equally spaced nodes lying on a torus 
embedded in $\s^3$, see Section \ref{s5}.  
If we define a unit $2$-vector $\bu_i$ from (\ref{e16}) according to
$\bx_i=\frac{1}{\sqrt{3}}(\sqrt{2}\bu_i,1)$ then 
$\sum_i\bu_i=0$ and so $\bu_i$ can be regarded as a
splay state as described in \cite{BS2021}.

Directly from (\ref{e16}) and the definition (\ref{f9}), together
with well-known trigonometric sums derived in \ref{A}, see 
in particular (\ref{a2}), we obtain $\av=\frac{1}{\sqrt{3}}(0,0,1)$.
The vectors (\ref{e16}) also satisfy, as proved in \ref{B}:
\beq
\label{e17}
\frac{1}{N^2}\sum_{j,k=1}^N\sgn_{ijk}(\bx_j\times\bx_k)
=
\frac{2}{N\sqrt{3}}\, \cot\frac{\pi}{N}\; \bx_i,
\eeq
for all $i$, which verifies the relation (\ref{e12}) for $d=3$,  with
$\lambda=\frac{2N}{\sqrt{3}}\cot\frac{\pi}{N}$. From (\ref{e17}) we
obtain
\[
\frac{1}{N^2}\sum_{j,k=1}^N\sgn_{ijk}(\bx_i\centerdot\bx_j\times\bx_k)
=
\frac{2}{N\sqrt{3}}\, \cot\frac{\pi}{N},
\]
from which $\sum_{j,k=1}^N\sgn_{ijk}\,
\left[\bx_j\times\bx_k-\bx_i\,(\bx_i\centerdot\bx_j\times\bx_k)\right]=0$.
Hence (\ref{e16}) is an exact static solution of (\ref{e13}), which numerically
we find to be stable for $\kappa_3>0$, while for $\kappa_3<0$ 
its negative is stable.


\subsection{3-body systems with distributed frequencies\label{ss32}}

Consider next the system (\ref{e13}) for distributed frequency
vectors $\bw_i$.  It is well-known for the $2$-body system (\ref{e2})
that in this case phase-locked
synchronization, in the sense that $r(t)=\|\av(t)\|$ approaches a 
constant value,  does not occur exactly for $d=3$.  
Rather, $r$ varies between
narrow limits which decrease as $\kappa_2\to\infty$, which has
been termed
practical synchronization \cite{CH2014,CCH2014}, i.e.\  
the particle positions do not shrink to a single point, but 
are confined to a small region with a diameter inversely proportional 
to $\kappa_2$ \cite{CH2014}.  This phenomenon of practical 
synchronization also appears for (\ref{e13}), since we find numerically
that $r(t)$ again varies asymptotically between narrow limits which
decrease as $|\kappa_3|$ increases. Ring synchronization, however, is still
clearly evident, as in the example in Fig.\ \ref{fig1}(b), which shows
an asymptotic configuration for $N=40$ for which the average value 
for $r$ is close to $\frac{1}{\sqrt{3}}$. The frequency vectors $\bw_i$ 
in this example
have random entries with approximate unit length, and $\kappa_3=20$.
Following the initial transient, the configuration rotates on 
$\s^2$ with small variations in the relative positions of the nodes.
For small values of $|\kappa_3|$ the particle  motion is asynchronous,
suggesting that there is a critical value $\kappa_c$ such that practical 
synchronization occurs only for $|\kappa_3|>\kappa_c>0$.


\subsection{An exact solution for $N=3$\label{ss43}}

The special case $N=3$ of (\ref{e13}) may be solved exactly for the rotational
invariants in terms of
elliptic functions, and provides insight into the rate of synchronization
and the stability of the synchronized configurations. 
In particular, we show why the fixed point corresponding to complete
synchronization is unstable, whereas the ring synchronized configuration,
corresponding to the $N=3$ case of (\ref{f12}),  is stable. 
With $\bw_i=0$ and $\kappa_3/N^2=1$, (\ref{e13}) reduces to:
\bea
\label{e19}
\dot{\bx_1}
&=&
2\left[\bx_2\times\bx_3-\bx_1\;(\bx_1\centerdot\bx_2\times\bx_3)\right]
\\
\nonumber
\dot{\bx_2}
&=&
2\left[\bx_3\times\bx_1-\bx_2\;(\bx_2\centerdot\bx_3\times\bx_1)\right]
\\
\nonumber
\dot{\bx_3}
&=&
2\left[\bx_1\times\bx_2-\bx_3\;(\bx_3\centerdot\bx_1\times\bx_2)\right],
\eea
hence
\[
\frac{d}{dt}(\bx_1\centerdot\bx_2)=
\bx_1\centerdot\dot{\bx_2}+\dot{\bx_1}\centerdot\bx_2
=
-4\bx_1\centerdot\bx_2\;(\bx_1\centerdot\bx_2\times\bx_3),
\]
together with the cyclic permutations.
Denote $x_{ij}=\bx_i\centerdot\bx_{j}, x_{ijk}=\bx_i\centerdot\bx_j\times\bx_k$,
 then we have
$\dot{x}_{12}=-4 x_{12} x_{123}, \dot{x}_{23}=-4 x_{23} x_{123}$ and
$\dot{x}_{13}=-4 x_{13} x_{123}$. Hence the ratios 
$x_{23}/x_{12}$ and $x_{13}/x_{12}$ are constants of the motion.
Let us choose $u=x_{12}$ to be the independent variable,  then
$x_{23}=c_1 u, x_{13}=c_2 u$ for constants $c_1,c_2$ which are fixed 
by the initial values $\bx_i^0$, i.e.\ we have 
$c_1=x_{23}^0/x_{12}^0, c_2=x_{13}^0/x_{12}^0$, 
assuming that $u_0=x_{12}^0=\bx_1^0\centerdot\bx_2^0$
is nonzero. The constants $c_1,c_2$ can take any value, positive 
or negative. 

Consider now the fixed points of the system (\ref{e19}). These
satisfy
$\bx_1\times\bx_2 =\bx_3\,x_{123}$ and its cyclic
permutations, which implies $x_{12} x_{123}=x_{13} x_{123}=x_{23} x_{123}=0$.
Either $x_{123}=0$, in which case all cross products are zero,
$\bx_i\times\bx_j=0$, or $x_{123}\ne0$ which implies
that the three vectors $\bx_i/x_{123}$ form an orthonormal set. 
In the first case, $\bx_1,\bx_2,\bx_3$ are either parallel
or antiparallel, with four possible choices of relative sign, and so the 
three nodes 
are located at either a single point, or at two opposite points on the 
unit circle.  In either case we have $|c_1|=|c_2|=1$, 
and since $c_1,c_2$ are constants of the motion, these fixed points 
exist for the system (\ref{e19}) only if the initial values $\bx_i^0$ 
are consistent with this,
i.e.\ only if  $|x_{12}^0|=|x_{13}^0|=|x_{23}^0|=1$.
These fixed points are therefore unstable, since under any perturbation
of the initial values these steady state solutions cannot be attained 
as the system evolves.

For the second case, in which $x_{123}\ne0$, we must have $|x_{123}|=1$ and
so $\{\bx_1,\bx_2,\bx_3\}$ forms an orthonormal set, with an
arbitrary orientation with respect to the axes, 
satisfying $x_{12}=x_{13}=x_{23}=0$.
Let us show that (\ref{e19}) synchronizes to these points from all
initial values, except for the unstable fixed points.
Firstly, we express $x_{123}^2$ in terms 
of $x_{12},x_{23},x_{13}$ by means of a well-known identity obtained as follows:
define the $3\times3$ matrix $M=(\bx_1,\bx_2,\bx_3)$ then
$(M^{\tp}M)_{ij}=x_{ij}$, and so  we obtain
$(\det M)^2=\det M^{\tp}\det M =\det (x_{ij})$, which is the Gram determinant, 
hence $x_{123}^2=
1-x_{12}^2-x_{13}^2-x_{23}^2+2x_{12}x_{13}x_{23}$.
Define the cubic polynomial
\beq
\label{e20}
p(u)=1-(1+c_1^2+c_2^2)u^2+2c_1c_2 u^3,
\eeq
then $x_{123}^2=p(u)$. From $\dot{u}=-4u\,x_{123}$ we obtain
$\dot{u}^2=16u^2 x_{123}^2=16u^2p(u)=2V(u)$, where we have defined the
potential $V(u)=8u^2p(u)$ as a fifth order polynomial in $u$.
The equation $\dot{u}^2=2V(u)$ can be solved for $u$ with 
the initial value $u(0)=u_0$,
however, in order to avoid taking the square root with
an ambiguous sign, it is preferable from a numerical perspective 
to solve the equivalent second order equation
$\ddot{u}=V'(u)$ with $u(0)=u_0, \dot{u}(0)=-4u_0x_{123}^0$. 

Let us determine the properties of $p$ and hence of $V$. We have
$p(0)=1, p(1)=-(c_1-c_2)^2$ and $p(-1)=-(c_1+c_2)^2$,
hence $p$ has a real root in $(0,1]$, denoted $r_+$, and a second real root
in $[-1,0)$, denoted $r_-$.  The third root $r_3=-1/(2c_1c_2r_-r_+)$ 
is therefore also real, and is positive or negative according to the sign
of $c_1 c_2$, and lies outside the interval $(-1,1)$. The possible
values of $u(t)$ for all $t>0$ are restricted by the condition
$p(u)=x_{123}^2\geqslant0$, which implies that
$-1\leqslant r_-\leqslant u(t)\leqslant r_+\leqslant1$ for all $t>0$.
We factorize $p$ according to
$p(u)= 2c_1c_2(r_+-u)(u-r_-)(r_3-u)$, then by integrating
$\dot{u}=\pm4u\sqrt{2c_1c_2(r_+-u)(u-r_-)(r_3-u)}$
we can obtain $u$ as an explicit elliptic integral of the third kind, 
see for example the expression in
Gradshteyn and Ryzhik  \cite{GR2014}, Section 3.137, item (3.) with $r=0$.

\begin{figure}[!ht]
\begin{center}
\includegraphics[width=0.48\columnwidth,trim={10 10 180 2},clip]{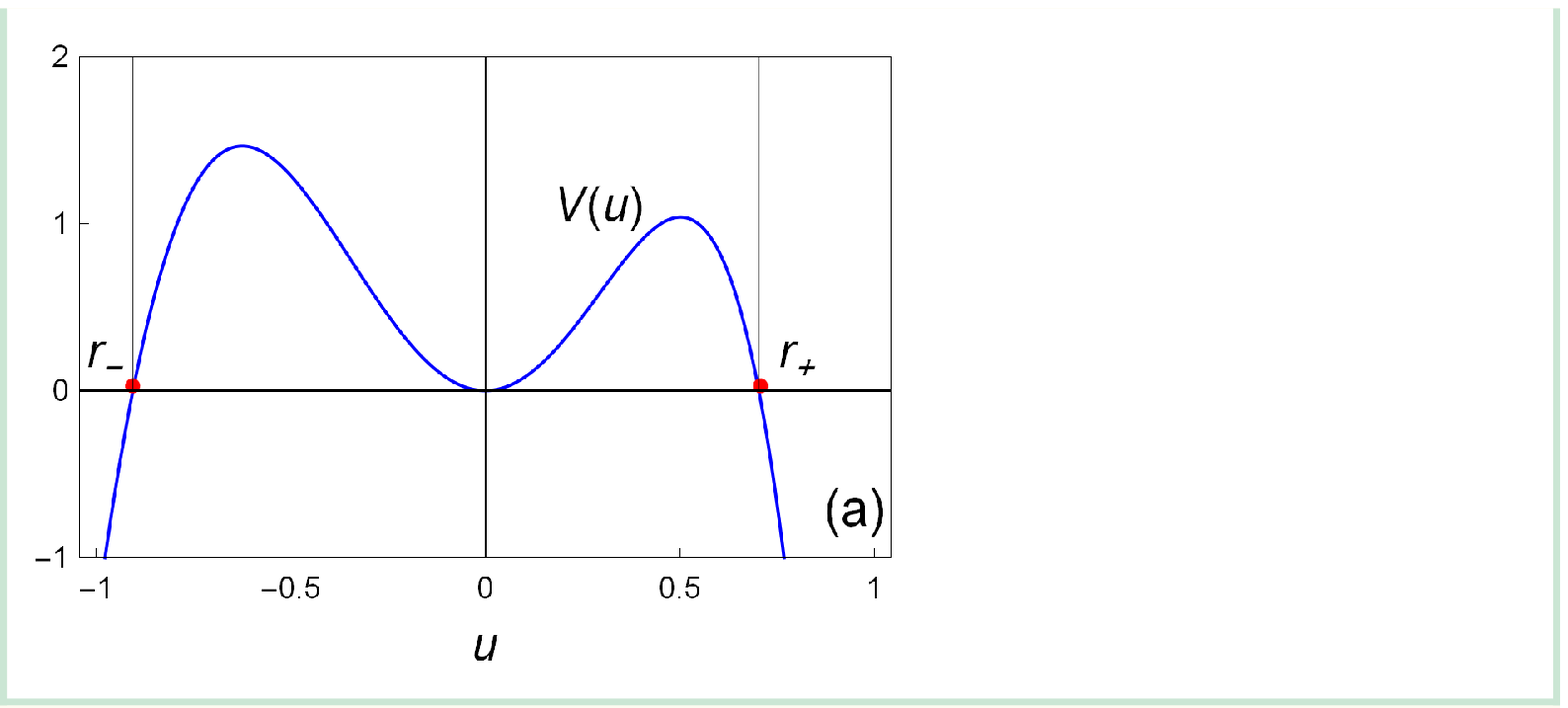}
\includegraphics[width=0.48\columnwidth,trim={2 10 190 2},clip]{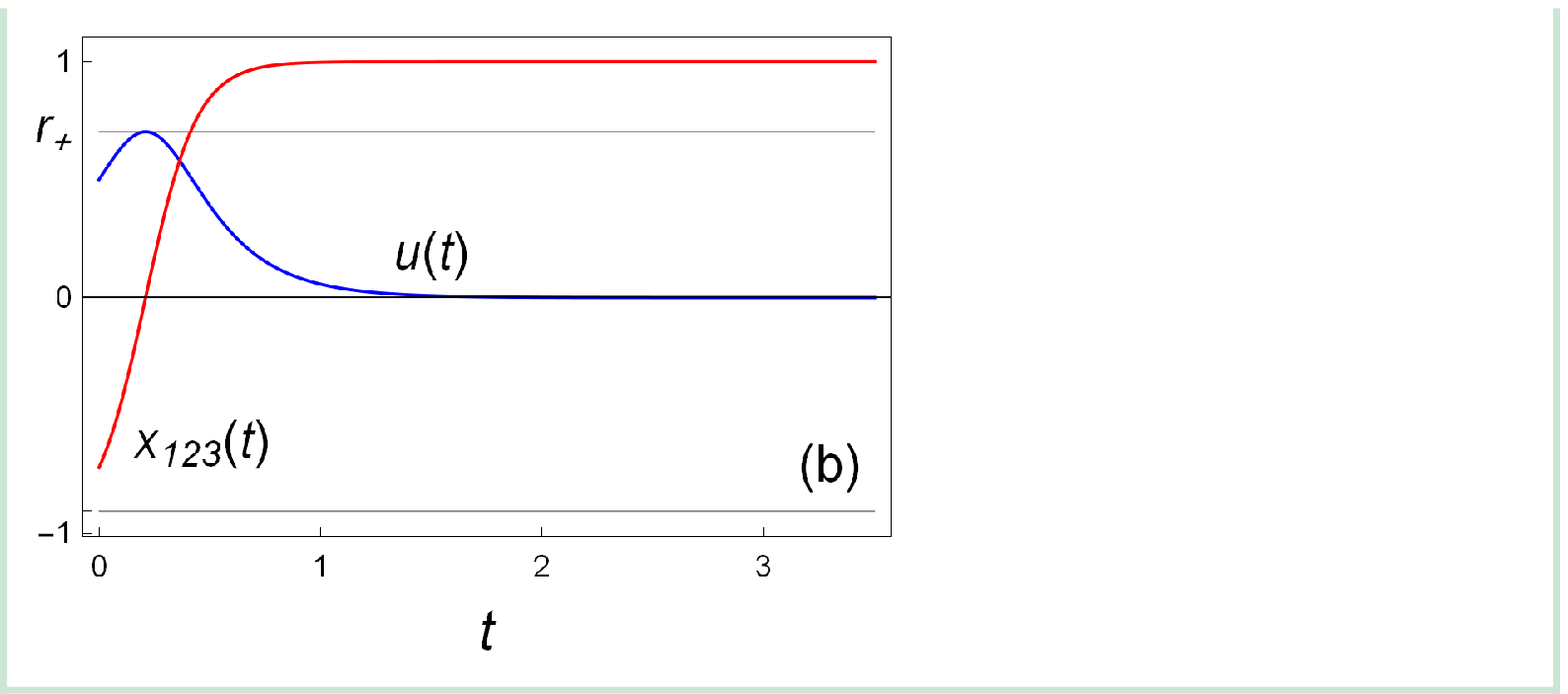}
\end{center}
\caption{
(a) The potential  $V(u)=8u^2p(u)$ for  $c_1=-\frac{3}{4},c_2=\frac{1}{3}$
showing the endpoints $r_{\pm}$ (in red) and the local minimum at $u=0$; (b)
the solutions $u(t)$ (blue) and $x_{123}(t)$ (red) 
for $u_0=\half$, showing $u$ approaching the stable fixed point at $u=0$.
}
\label{fig2}
\end{figure}

We can determine how the solution behaves, however, without an explicit
expression for $u$. There is a mechanical analogy with a particle moving
under the influence of the potential $V$ with the Lagrangian
$L=T-V=\half {\dot{u}}^2-V(u)$, with the corresponding equation of motion
$\ddot{u}=V'(u)$, but with boundary conditions such that the first integral 
$\dot{u}^2=2V(u)$ holds.  The potential $V(u)=8u^2p(u)$ has zeroes 
in $[-1,1]$ at $u=0$ and at $u=r_{\pm}$, and 
for all initial values, except $u_0= r_{\pm}$, 
the particle settles into the stable minimum of  $V$ at $u=0$. 
An example of $V$ is shown in Fig.\ \ref{fig2}(a)
for the parameters $c_1=-\frac{3}{4},c_2=\frac{1}{3}$, corresponding
to a specific choice for the initial values $\bx_1^0,\bx_2^0,\bx_3^0$.  
The motion of a particle located at $u$ is restricted to lie between 
the limits $r_{\pm}$, the roots of $p$ as shown in Fig.\ \ref{fig2}(a), 
and are given explicitly for this example by $r_-=-0.905, r_+=0.703$. 
These points are static solutions of $\dot{u}^2=16u^2p(u)$, but do not
correspond  to any fixed point solutions of the full system (\ref{e19}),
unless $|c_1|=|c_2|=1$. 
They are, in any case, unstable fixed points of the combined 
equations for $u$ and $x_{123}$, for any $c_1,c_2$. From 
$x_{123}^2=p(u)$ and $\dot{u}=-4 u \,x_{123}$ we obtain 
$\dot{x}_{123}=-2u\, p'(u)$, the right-hand side of which is non-negative
for all $u$, and so $x_{123}$ is an increasing function of $t$.
In particular, $\dot{x}_{123}$ is nonzero at either endpoint 
$r_{\pm}$, and so these endpoints are
unstable static solutions of $\dot{u}^2=16u^2p(u)$. 
If a particle initially moves to one of these endpoints, it
simply bounces off the barrier imposed by the condition
$r_-\leqslant u\leqslant r_+$, and then approaches the stable equilibrium
point at $u=0$. For the potential $V$ in Fig.\ \ref{fig2}(a), we have 
plotted  the explicit solutions $u(t),x_{123}(t)$ for the initial
values $u_0=\half, x_{123}^0=-\sqrt{p(u_0)}$ in Fig.\ \ref{fig2}(b),
showing $x_{123}(t)$ (in red) as an increasing function of $t$ with
$x_{123}(t)\to1$, and $u$ bouncing off the point at $u=r_+$ and
then approaching the asymptotic limit of zero, which is
the stable minimum of $V$. The fact that $x_{123}$ is an increasing
function of time is a general property of the gradient
formulation of the system, since by regarding
the potential $\V_d$ as a function of time, we have
$\dot{\V}_d=\sum_i\nabla_i\V_d\centerdot\dot{\bx}_i=\sum_i\|\nabla_i\V_d\|^2>0$.

We deduce that for all initial values $\bx_1^0,\bx_2^0,\bx_3^0$,
except for the unstable fixed points with $
|c_1|=|c_2|=1$, the system synchronizes to  $x_{12}=x_{13}=x_{23}=0$.
The vectors $\bx_1,\bx_2,\bx_3$ therefore
form an orthonormal set which by means of a rotation we can align with the
$X,Y,Z$ axes (up to a left-right orientation), and so the nodes lie in the
plane $x+y+z=1$ with the unit normal $\bn=\frac{1}{\sqrt{3}}(1,1,1)$.
As before, we obtain $r_{\infty}=\frac{1}{\sqrt{3}}$. 
The relation (\ref{e17}) is satisfied with 
$\lambda=\frac{2N}{\sqrt{3}}\cot\frac{\pi}{N}=2$.

Although
we have determined the rotational 
invariants $x_{12},x_{13},x_{23},x_{123}$
as functions of $t$, we omit the further step of solving 
(\ref{e19}) explicitly for $\bx_1(t),\bx_2(t),\bx_3(t)$,
which requires us to choose a suitable basis, since 
synchronization of (\ref{e19}) follows from the properties of the 
rotational invariants and the  fixed points.

\section{Combined $2$- and 3-body interactions on the $2$-sphere\label{sec4}}

Of particular interest with all higher-order interactions is how they compare
with the well-studied $2$-body interactions, and whether one is dominant
in some sense. 
The $2$-body and 3-body systems for $d=3$, given by (\ref{e2}) and (\ref{e13})
respectively, are ideal  for such an investigation since,
separately, they each synchronize although under
different conditions, and generally with incompatible properties.
The $2$-body system (\ref{e2}), with identical frequency matrices
$\Omega_i=\Omega$ and global coupling $a_{ij}=1$, completely synchronizes
for $\kappa_2>0$, with the order parameter $r$ evolving
exponentially quickly to the asymptotic value $r_{\infty}=1$ 
\cite{CCH2014,CH2014}. 
Evidently all nodes strongly attract each other from all initial values.
For $\kappa_2<0$, however, the system
evolves to a final state with $r_{\infty}=0$, in which the particles 
are distributed over $\s^2$ \cite{CH2014}, and so for this case
the interactions are repulsive, a property well-known also
for the Kuramoto model with $d=2$.

By contrast, the 3-body system (\ref{e13}) with identical frequencies
$\bw_i=\bw$, synchronizes for any $\kappa_3$ from generic initial
values to a final configuration with $r_{\infty}=\frac{1}{\sqrt{3}}$,
in which the particles are equally separated as shown in Figure \ref{fig1}
(left), which we refer to as ring synchronization.
For combined $2$-body and 3-body interactions the question arises, therefore, 
if the initial particles are clustered together in a tight bunch,
do they synchronize to a point under the influence of the attractive 
$2$-body interactions, or do they separate according to the 3-body forces
to form a ring on $\s^2$? As will be shown, the system
synchronizes for any coupling coefficients, whether positive or negative,
and the 3-body forces prevail unless
the $2$-body coupling is positive and sufficiently strong by comparison to 
the 3-body coupling. The specific system
we consider, the $d=3$ case of (\ref{g13}), is given by:
\beq
\fl
\label{f20}
\dot{\bx_i}
=
\kappa_2\av-\kappa_2\,\bx_i(\bx_i\centerdot\av)
+
\frac{\kappa_3}{N^2}\sum_{j,k=1}^N \sgn_{ijk} \left[\bx_j\times\bx_k
-\bx_i\;(\bx_i\centerdot\bx_j\times\bx_k)\right],
\eeq
and properties of the system depend on the coupling constants 
only through the ratio $\kappa_2/\kappa_3$.

We find that (\ref{f20}) achieves ring synchronization for all 
positive and negative values of $\kappa_2,\kappa_3$, unless $\kappa_2/\kappa_3$
exceeds a critical value, specifically, (\ref{f20}) 
completely synchronizes only for 
$\kappa_2/|\kappa_3|> \frac{2}{N}\cot\frac{\pi}{N}$, otherwise the 
3-body forces prevail. The system ring synchronizes for all $\kappa_2<0$ 
for which the $2$-body forces are repulsive, as we show next.


\subsection{Steady state solutions on $\s^2$\label{ss51}}

Numerically we find that solutions of (\ref{f20}), for generic initial values, 
approach a static configuration. 
The vectors $\bx_i=\pm \,\bu$ are fixed points of (\ref{f20}), where the sign 
can depend on $i$, and indeed for 2-body interactions with $\kappa_3=0$ 
and $\kappa_2>0$ we obtain 
complete synchronization with plus signs for all $i$ \cite{CH2014}, and
then $\bx_i=\bu$ is a stable fixed point.  We can also obtain
bipolar synchronization, in which the signs vary, if we include
multiplicative factors of any sign in the $2$-body interactions
\cite{ML2014,HKLN2020}.

The following configuration,
which generalizes (\ref{e16}), is a steady state solution of the
combined system (\ref{f20}):
\beq
\label{e21}
\bx_i=
r_{\infty}\left(\alpha\cos\frac{2i\pi}{N},
\alpha\sin\frac{2i\pi}{N},1\right),\quad i=1, \dots N,
\eeq
where $r_{\infty},\alpha$, to be determined, are parameters 
satisfying $r_{\infty}^2(1+\alpha^2)=1$, which ensures that 
$\bx_i$ is a unit vector. As before, we have used the rotational covariance of 
(\ref{f20}) to rotate any planar static configuration to lie in the
plane $z=r_{\infty}$, together with an $\mathrm{SO}(2)$ rotation
about the $Z$-axis so that 
$\bx_N=r_{\infty}\left(\alpha,0,1\right)$ is aligned along the
$X$-axis. The rotational invariant corresponding to
(\ref{e21}) is
\beq
\label{x20}
\bx_i\centerdot\bx_{j}
=
r_{\infty}^2\left[1+\alpha^2\cos\frac{2(i-j)\pi}{N}\right],
\eeq
and so again the nodes are equally spaced. We determine
$r_{\infty}$, and hence $\alpha$, as a function of $\kappa_2/\kappa_3$ by 
requiring that the steady state (\ref{e21}) should satisfy (\ref{f20}).

Let us firstly verify that the parameter $r_{\infty}$ in (\ref{e21})
corresponds to $\|\av\|$. We have
\[
\av
=
\frac{1}{N}\sum_{j=1}^N\bx_j
=
r_{\infty}\left(\frac{\alpha}{N}\sum_{j=1}^N\cos\frac{2j\pi}{N},
\frac{\alpha}{N}\sum_{j=1}^N\sin\frac{2j\pi}{N},1\right)
=
r_{\infty}\left(0,0,1\right),
\]
where we have used (\ref{a2}),
and hence $\|\av\|=r_{\infty}$ as required. Also, from (\ref{e21}):
\beq
\label{e22}
\av-\bx_i(\av\centerdot\bx_i)
=
r_{\infty}\left(0,0,1\right)-r_{\infty}^2\bx_i,
\eeq
and 
\[
\fl
\bx_j\times\bx_k
=
r_{\infty}^2\alpha
\left(
\sin\frac{2j\pi}{N}-
\sin\frac{2 k\pi}{N},
-\cos\frac{2j\pi}{N}+
\cos\frac{2k\pi}{N},
-\alpha\sin\frac{2(j-k)\pi}{N}\right).
\]
By means of the summation formulas in \ref{B} we obtain:
\beq
\label{e23}
\fl
\frac{1}{N}\sum_{j,k=1}^N \sgn_{ijk}\; (\bx_j\times\bx_k)
=
2r_{\infty} \cot\frac{\pi}{N}
\;\bx_i
+
\frac{(1-3r_{\infty}^2)}{r_{\infty}}
\cot\frac{\pi}{N}\,\av,
\eeq
from which
\beq
\label{e24}
\frac{1}{N}\sum_{j,k=1}^N \sgn_{ijk}\;(\bx_i\centerdot\bx_j\times\bx_k)
=
3r_{\infty}(1-r_{\infty}^2)\cot\frac{\pi}{N},
\eeq
for all $i$. 

Evidently  (\ref{e23}) corresponds to the general formula (\ref{g14}) 
with $\lambda_1=2Nr_{\infty} \cot\frac{\pi}{N}$ and 
$\lambda_2=N\cot\frac{\pi}{N}(1-3r_{\infty}^2)/r_{\infty}$.
Hence (\ref{f20}) is satisfied provided that (\ref{f15}) holds, i.e.\
provided that
$\kappa_2 r_{\infty} +\kappa_3(1-3 r_{\infty}^2)\frac{1}{N}
\cot\frac{\pi}{N}=0$. We have established therefore that
(\ref{e21}) is a steady state solution of (\ref{f20}), which we find
numerically to be stable for $\kappa_3>0$, under conditions to be determined.
For $\kappa_3<0$ the stable solution is found by means of the parity change 
$\bx_i\to-\bx_i$ in (\ref{f20}),
by reversing the sign of $\kappa_3$. The condition therefore that 
(\ref{e21}), or its negative, be a stable steady state for (\ref{f20})
is:
\beq
\label{e25}
\kappa_2 r_{\infty} +\frac{|\kappa_3|}{N}(1-3 r_{\infty}^2) \cot\frac{\pi}{N}=0,
\eeq
which leads to the following explicit expression for $r_{\infty}$ as a function
of $\kappa_2/\kappa_3$:
\beq
\label{e26}
r_{\infty}
=
\frac{\kappa_2  N\tan\frac{\pi}{N}}{6|\kappa_3|}
+\frac{1}{6}\sqrt{\frac{\kappa_2^2 N^2 \tan^2\frac{\pi}{N}}{\kappa_3^2}+12}.
\eeq
Evidently $r_{\infty}$ is positive for any sign combination of 
$\kappa_2,\kappa_3$, and for $\kappa_2=0$ we regain the 
value $r_{\infty}=\frac{1}{\sqrt{3}}$ as discussed in Section \ref{s4}.
If $\kappa_3=0$ the vector (\ref{e21}) is not a fixed point of the
$2$-body system, as (\ref{e22}) shows directly, unless
$r_{\infty}=1,\alpha=0$.


\subsection{Transition to complete synchronization\label{ss42}}

The formula (\ref{e26}) for $r_{\infty}$ violates the bound 
$r_{\infty}\leqslant1$ if $\kappa_2/|\kappa_3|$ is too large. 
The critical ratio is found by setting
$r_{\infty}=1$ in (\ref{e25}), hence we require
\beq
\label{e28}
\frac{\kappa_2}{|\kappa_3|}\leqslant\frac{2}{N}\cot\frac{\pi}{N},
\eeq
in order for the  static solution (\ref{e21}) of the combined
system (\ref{e20}) to exist. 

We can now determine the behaviour of the system as $\kappa_2/\kappa_3$
varies. As $\kappa_2/|\kappa_3|$ approaches the critical ratio 
from below,  it follows from (\ref{e26}) that 
$r_{\infty}\to1$, which means that the particles become more tightly
bunched, and so the synchronized ring shrinks in diameter and eventually 
reduces to a point at which equality holds in (\ref{e28}). 
Complete synchronization occurs
for all larger values of $\kappa_2/|\kappa_3|$, i.e.\ the 2-body forces prevail
through a continuous transition.
Otherwise, whenever $\kappa_2/|\kappa_3|$ is less than 
$\frac{2}{N}\cot\frac{\pi}{N}$, ring synchronization prevails, 
in particular as $\kappa_2$ decreases to large negative values,
we have  $r_{\infty}\to0$ as
is evident from the formula (\ref{e26}), and so the ring expands to a 
great circle on $\s^2$. In this case the 3-body forces overcome
the repulsive 2-body forces, however large, to maintain ring synchronization.

\begin{figure}[!ht]
\begin{center}
\includegraphics[width=0.49\columnwidth,trim={40 80 120 60},clip]{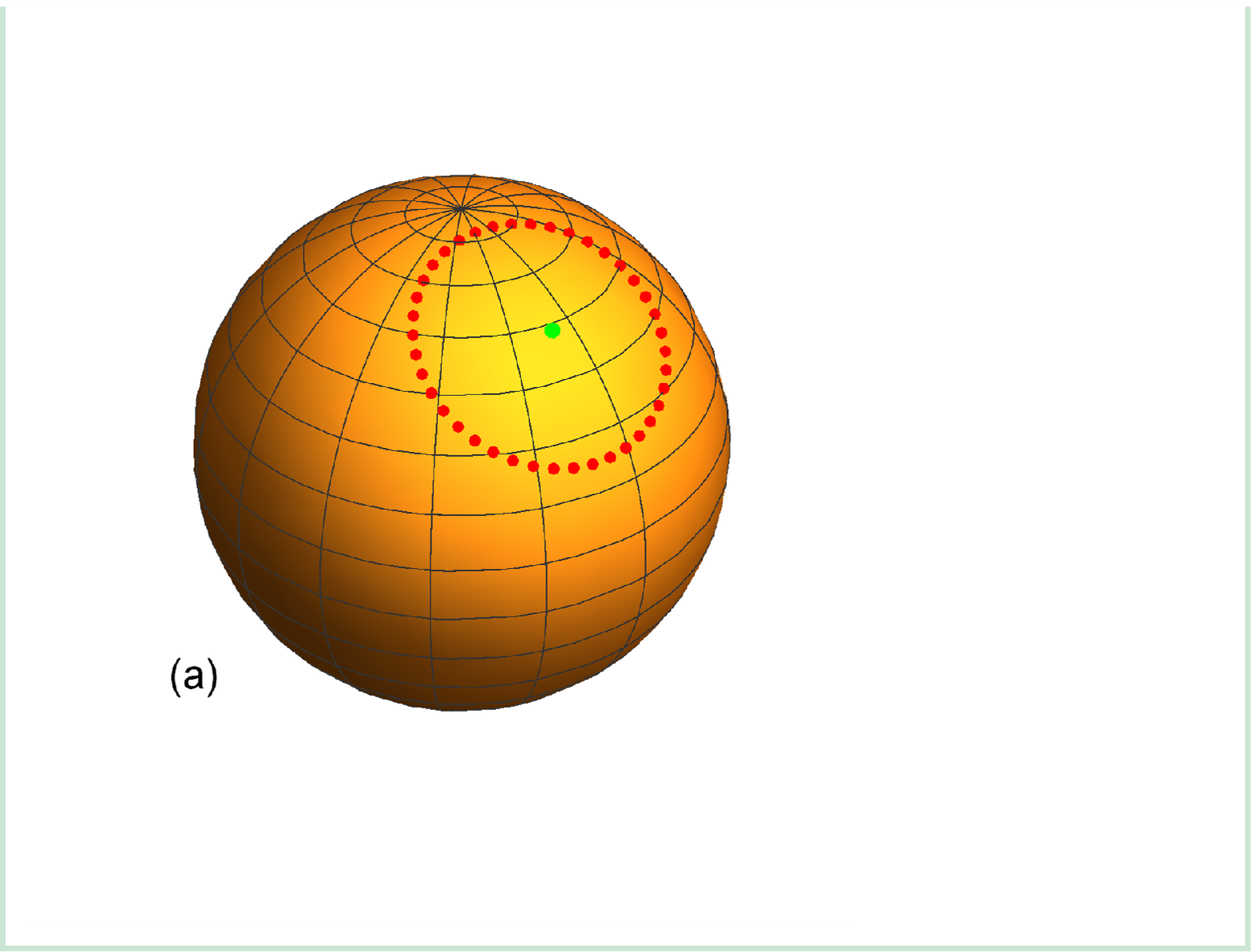}
\includegraphics[width=0.49\columnwidth,trim={40 80 120 60},clip]{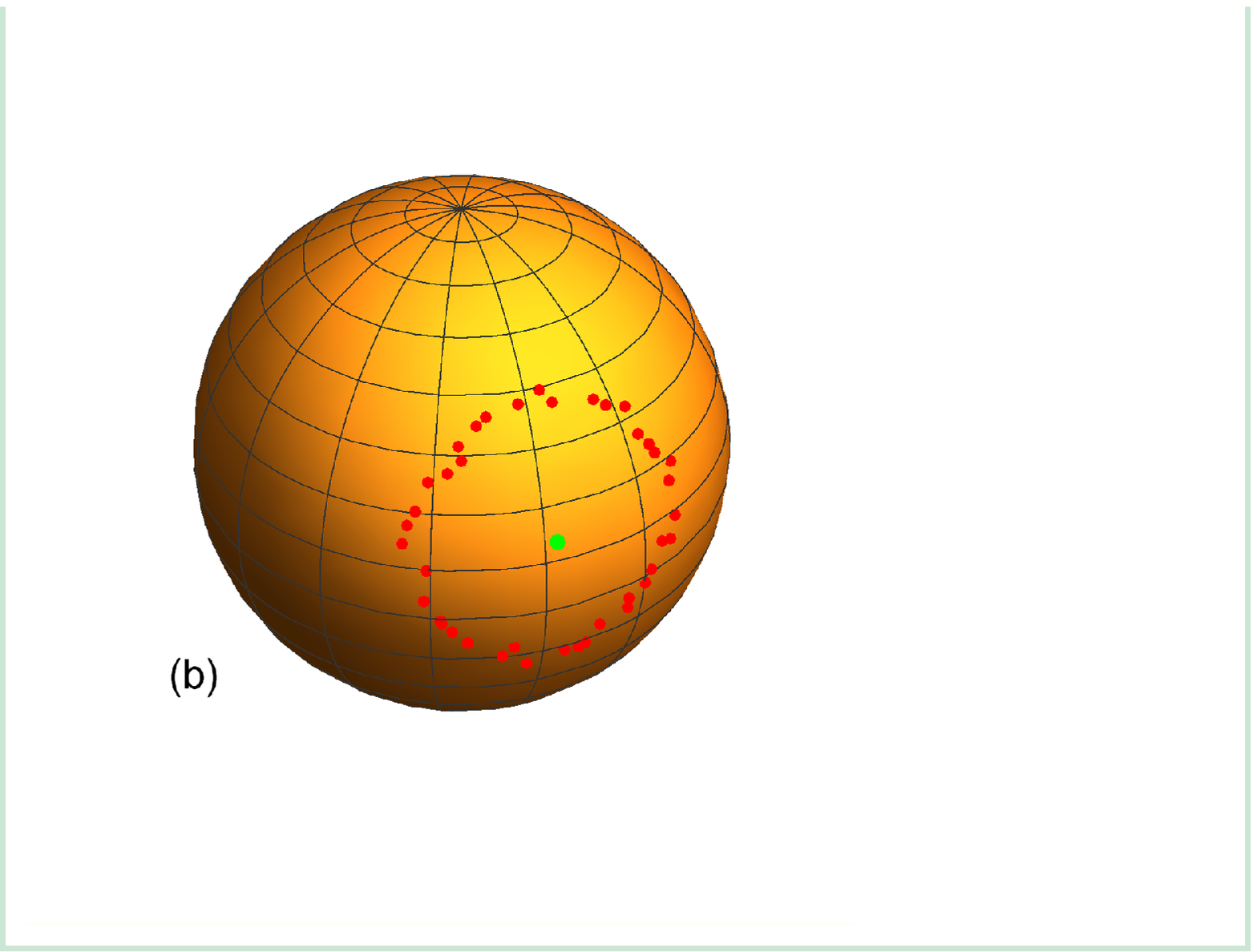}
\end{center}
\caption{
(a) An example of ring synchronization for the system (\ref{e13}) with $N=40,
\kappa_2=1, \kappa_3=2$, showing a ring of reduced diameter with 
$r_{\infty}\approx0.896$;
(b)
the same system but with distributed frequencies $\bw_i$ with 
$\|\bw_i\|<1/20$, showing approximate (practical) ring synchronization. 
}
\label{fig3}
\end{figure}

Numerical calculations confirm these properties. Fig.\ \ref{fig3}(a) shows
ring syn\-chronization for the system (\ref{f20}) with 
$N=40,\kappa_3=2,\kappa_2=1$, for which $\kappa_2/|\kappa_3|$ is less 
than the critical ratio $0.635$, and numerically we find $r_{\infty}=0.896$, 
in agreement with (\ref{e26}). Evidently, the attractive 2-body forces 
shrink the ring of synchronized nodes down to a smaller diameter,
compared to that in Fig.\ \ref{fig1}(a).
If we include frequency terms $\bw_i\times\bx_i$ on the right-hand side
of (\ref{f20}), then exact synchronization is replaced by practical
synchronization, in which the final configuration, now time-dependent,
resembles either the ring or the completely synchronized
configuration that occurs for $\bw_i=0$, depending on $\kappa_2/\kappa_3$. 
As an example,
in Fig.\ \ref{fig3}(b) we show the effect of nonzero frequency
vectors $\bw_i$ such that $\|\bw_i\|<1/20$, for the same parameters as in 
Fig.\ \ref{fig3}(a), where evidently the synchronized ring is slightly 
distorted by the small nonzero oscillations at each node.

Practical synchronization occurs even for $\kappa_2<0$, when the
$2$-body forces are repulsive, provided that $|\kappa_3|$ is sufficiently
large to overcome the natural oscillations. If we fix $\kappa_2=-1$ 
and choose the same frequencies $\bw_i$ as in Fig.\ \ref{fig3}(b), then
the system synchronizes with $r_{\infty}\approx0.366$ for $\kappa_3=2$,
and for any larger values, but not for $\kappa_3=1$, which suggests that 
there exists a critical value for $|\kappa_3|$, in order
for synchronization to occur.
Again, the 3-body forces enhance synchronization, since without
3-body interactions of sufficient strength the system would evolve
asynchronously for any $\kappa_2<0$.

\section{4-body interactions on the 3-sphere\label{s5}}

The $4$-body forces have properties which differ from
those  for $d=3$, consistent with previous observations that
features of even-dimensional systems on the sphere differ
from those for odd dimensions \cite{ML2009,Chandra2019}.
Whereas the $d=3$ system always synchronizes  with the same value
$r_{\infty}=\frac{1}{\sqrt{3}}$, now $r_{\infty}$ depends
on $N$, and although the nodes arrange themselves to lie equally spaced,
in this case on a torus imbedded in $\s^3$, the nodes do not form a closed
loop. This also occurs for $d=2$, see for example Fig.\ \ref{fig5}(a).


\subsection{The homogeneous system}

Consider firstly the homogeneous 4-body system 
\beq
\label{e27}
\dot{\bx_i}
=
\frac{\kappa_4}{N^3} \sum_{j,k,l=1}^N \sgn_{ijkl}
\left[\bv_{jkl}-\bx_i(\bx_{i}\centerdot\bv_{jkl})
\right],
\eeq
where the antisymmetric vectors $\bv_{jkl}$ are defined by (\ref{f7}).  We can 
reverse the sign of $\kappa_4$ by means of an element $R\in\mathrm{O}(4)$
such that $\det R=-1$, and by rescaling we set $\kappa_4/N^3=1$.

Numerically, we find that the system (\ref{e27}) synchronizes
from generic initial values to a final static configuration in which
the scalars $\bn\centerdot\bx_i=n_i$, where
$\bn=\av/\|\av\|$, depend on $i$, always with $n_i>0$. Hence the 
final configuration is confined to one hemisphere of $\s^3$ but,
unlike the $d=3$ case,  does not lie in a hyperplane which
intersects $\s^3$.
The following rotationally invariant formula for the static 
final configuration holds for all $i,j$:
\beq
\label{e29}
\bx_i\centerdot\bx_{j}
=
\half\cos\frac{\pi(i-j)}{N}
+
\half\cos\frac{3\pi(i-j)}{N},
\eeq
which implies that the nodes are equally spaced, i.e.\
$\|\bx_i-\bx_{i+1}\|$ is independent of $i$ for $i=1,\dots N-1$.
The symmetry $\bx_i\centerdot\bx_{N}=-\bx_{N-i}\centerdot\bx_{N}$ also
follows. We deduce the following expression for $\bx_i$, up to a constant
rotation:
\beq
\label{e30}
\bx_i=\frac{1}{\sqrt{2}}\left(\cos\frac{\pi i}{N},
\sin\frac{\pi i}{N},\cos\frac{3\pi i}{N},\sin\frac{3\pi i}{N}
\right).
\eeq 
These $N$ nodes do not form a closed loop since for $i=0$ we have
$\bx_0=(1,0,1,0)/\sqrt{2}$ which equals $-\bx_N$, rather than $\bx_N$.
The configuration (\ref{e30}) consists of $N$ equally spaced
nodes  lying on a torus imbedded in $\s^3$, with the first 
two components restricted to a semicircle, while the last two
components trace out a full circle plus a semicircle for $i=1,\dots N$.

The expression (\ref{e30}) satisfies (\ref{e27}) exactly, as a consequence
of the following relation, which is the $d=4$ case of (\ref{e12}):
\beq
\label{e35}
\sum_{j,k,l=1}^N \sgn_{ijkl}\,\bv_{jkl}
=\frac{3N}{2}\cot\frac{3\pi}{2 N}\cot\frac{\pi}{2 N}\; \bx_i,
\eeq
which we prove in \ref{D}.  
From  (\ref{e30}), with the help of (\ref{a3}), we find:
\[
\av
=
\frac{1}{N\sqrt{2}}\left(-1,\cot\frac{\pi}{2N} , -1,\cot\frac{3\pi}{2N} \right),
\]
from which we obtain
\beq
\label{g37}
r_{\infty}^2
=
\frac{1}{2N^2}\left(\frac{1}{\sin^2\frac{\pi}{2N}} 
+\frac{1}{\sin^2\frac{3\pi}{2N}} \right).
\eeq
Evidently, $r_{\infty}$ is
a function of $N$, decreasing from its maximum value 
$r_{\infty}=\half$ at $N=4$,  to a minimum which is attained
as $N\to\infty$ for which $r_{\infty}\to 2\sqrt{5}/(3\pi)\approx 0.4745$.

Although the nodes (\ref{e30}) lie on $\s^3$, we can visualize
the configuration
by omitting one of the four components, then normalizing the remaining
components to form a unit 3-vector on $\s^2$.  We plot
such a configuration in Fig.\ \ref{fig4}(a) for $N=80$, in which the second
component has been dropped, showing equal spacing
with endpoints that are diametrically opposite on $\s^2$.
Another way of visualizing the evolving system is by means of the
Hopf fibration in which $\s^3$ is mapped to $\s^2$ according to
\beq
\label{f37}
(x,y,z,w)\to \Big(2(xz-yw),2(xw+yz),x^2+y^2-z^2-w^2\Big).
\eeq
In general, this maps circles on $\s^3$ to points on $\s^2$, but here 
distinct points on $\s^3$ are mapped one-to-one or two-to-one to points
on $\s^2$, which enables us to visualize the evolving system as it 
synchronizes.
The Hopf fibration and therefore the final configuration, however, 
does not respect the  
$\mathrm{SO}(4)$ covariance of the system (\ref{e27}) and so the final nodes
are generally not arranged in a symmetrical configuration. Equal spacing
of the nodes, for example, which follows from the rotationally invariant
expression (\ref{e29}), is not evident. However, if we rotate
the final configuration to obtain (\ref{e30}),   then
under the Hopf fibration (\ref{f37}) we find that
$\bx_i\to\left(\cos\frac{4 \pi i}{N},\sin\frac{4 \pi i}{N},0\right)$, and
so for odd $N$ the nodes are mapped to $N$ equally spaced distinct
points on the equator of $\s^2$, and to $N/2$ distinct points for even $N$.


\subsection{Combined pairwise and 4-body interactions on 
the 3-sphere\label{ss52}}

The combined $2$- and 4-body system is given by (\ref{g13}),
namely:
\beq
\fl
\dot{\bx}_i
=
\kappa_2\av-\kappa_2\,\bx_i\,(\bx_i\centerdot\av)+
\frac{\kappa_4}{N^3} \sum_{j,k,l=1}^N \sgn_{ijkl}
\left[\bv_{jkl}-\bx_i\,(\bx_{i}\centerdot\bv_{jkl})
\right]
\label{e36}.
\eeq
Numerically this system synchronizes in a way similar to the $d=3$ system
(\ref{f20}), in that the 4-body forces dominate so that the final
configuration is similar to that with $\kappa_2=0$, unless
the ratio $\kappa_2/\kappa_4$ is sufficiently large, in which case
complete synchronization occurs. We find that the 
expression (\ref{e29}) generalizes to:
\beq
\label{e37}
\bx_i\centerdot\bx_{j}
=
\cos^2\theta\cos\frac{\alpha\pi (i-j)}{N}
+
\sin^2\theta\cos\frac{3 \beta \pi (i-j)}{N},
\eeq
where $\alpha,\beta ,\theta$ are unknown parameters. This relation
can be verified numerically to high accuracy for suitably fitted parameters 
$\alpha,\beta ,\theta$, indicating that the form
(\ref{e37}) is exact, however there
is no simple functional relation between $\alpha$ and $\beta$.
For the homogeneous case (\ref{e30}) we have $\theta=\frac{\pi}{4}$ and 
$\alpha=\beta =1$, but
otherwise $\alpha,\beta$ satisfy nontrivial trigonometric 
relations that depend on $N$. Consistent with (\ref{e37}), we have
\beq
\label{e40}
\bx_i=
\left(\cos\theta\cos\frac{\alpha \pi i}{N},\cos\theta\sin\frac{\alpha \pi i}{N},
\sin\theta\cos\frac{3 \beta  \pi i}{N},\sin\theta\sin\frac{3 \beta  \pi i}{N}
\right),
\eeq
and the relations between $\alpha,\beta ,\theta$  are determined
by requiring that (\ref{g14}) be exactly satisfied 
for constants $\lambda_1,\lambda_2$ which depend on $\alpha,\beta ,\theta$. 
Numerically, we find indeed that (\ref{g14})
is satisfied in all cases, for constants $\lambda_1,\lambda_2$ which are
independent of $i$ for all initial values.
It is possible in principle to derive exact expressions for all 
quantities of interest by means of (\ref{e40}),
as we discuss briefly in \ref{D}, however the resulting equations are of
such complexity, due to the fact that $\alpha,\beta$ are not integers, 
that we confine ourselves here mainly to numerical observations.

One way of visualizing the configuration (\ref{e40})
is by means of the Hopf fibration (\ref{f37}), for which
\[
\bx_i\to
\left(\sin2\theta\cos\frac{i(\alpha+3\beta)\pi}{N},
\sin2\theta\sin\frac{i(\alpha+3\beta)\pi}{N}, \cos2\theta
  \right),
\]
showing that the nodes form a ring of equally spaced points 
around the 2-sphere in the plane $z=\cos2\theta$. The nodes become
more tightly spaced as $\alpha,\beta$ decrease, in which case $r_{\infty}$
increases.
We can determine the following expression for $r_{\infty}^2$ from
either (\ref{e37}) or (\ref{e40}):
\beq
\label{e42}
r_{\infty}^2
=
\frac{\cos^2\theta}{N^2}
\frac{\sin^2\frac{\pi \alpha}{2}}{\sin^2\frac{\pi\alpha}{2N}}
+
\frac{\sin^2\theta}{N^2}
\frac{\sin^2\frac{3\pi \beta}{2}}{\sin^2\frac{3\pi \beta}{2N}},
\eeq
which reduces to (\ref{g37}) for $\theta=\frac{\pi}{4},\alpha=\beta =1$.
We wish to determine the properties of the system as
the $2$-body forces increase in strength
relative to the 4-body forces. For $d=3$ we found that a continuous
transition occurs at a critical ratio, from ring synchronization to 
complete synchronization, but here for $d=4$ we find that this
transition is discontinuous, i.e.\ as $\kappa_2/\kappa_4$ increases
the system suddenly completely synchronizes at, and beyond, a critical ratio.
The order parameter $r_{\infty}^2$ as given in (\ref{e42})
attains the value unity, signifying complete synchronization, only
if both $\alpha,\beta\to0$, as can be deduced from (\ref{e42}), but
numerically we find that neither $\alpha$ nor $\beta$ approach zero as 
the transition point is approached, which is indicative of a discontinuity.

Other properties of the combined system (\ref{e36}) are similar to the 
$d=3$ case, for example the system synchronizes to a configuration satisfying 
(\ref{e37}) for all negative values of $\kappa_2$. If we fix $\kappa_4$ and
allow $\kappa_2$ to decrease to large negative values, $r_{\infty}$ decreases
accordingly, with the equal spacing of nodes in all cases 
maintained under the repulsive 2-body forces. Again, the 4-body forces 
prevail over the repulsive 2-body forces so as to enforce synchronization.


\section{Combined $5$-body and $2$-body interactions on $\s^4$\label{s6}}

The largest value of $d$ that we consider is
$d=5$ for the combined system (\ref{g13}), in order to confirm
that it has properties similar to the $d=3$ case. 
The main difference compared with $d=3$ is that as the 2-body 
forces increase in relative strength, the transition to a 
completely synchronized configuration is discontinuous. We have the equations:
\beq
\fl
\label{e43}
\dot{\bx}_i
=
\kappa_2\av-\kappa_2\,\bx_i\,(\bx_i\centerdot\av)
+
\frac{\kappa_5}{N^4}
\sum_{j,k,l,m=1}^N\sgn_{ijklm}\left[\bv_{jklm}
-
\bx_i\,(\bx_i\centerdot\bv_{jklm})\right],
\eeq
and again find numerically that
the system synchronizes for all generic initial values, either to
a configuration with equally spaced nodes, even for negative $\kappa_2$, 
or else to a point
if $\kappa_2/\kappa_5$ is sufficiently large. This holds for either
 sign of $\kappa_5$, which can be reversed by replacing $\bx_i\to-\bx_i$.
 

\subsection{The homogeneous 5-body system}

\begin{figure}[!ht]
\begin{center}
\includegraphics[width=0.49\columnwidth,trim={40 60 160 40},clip]{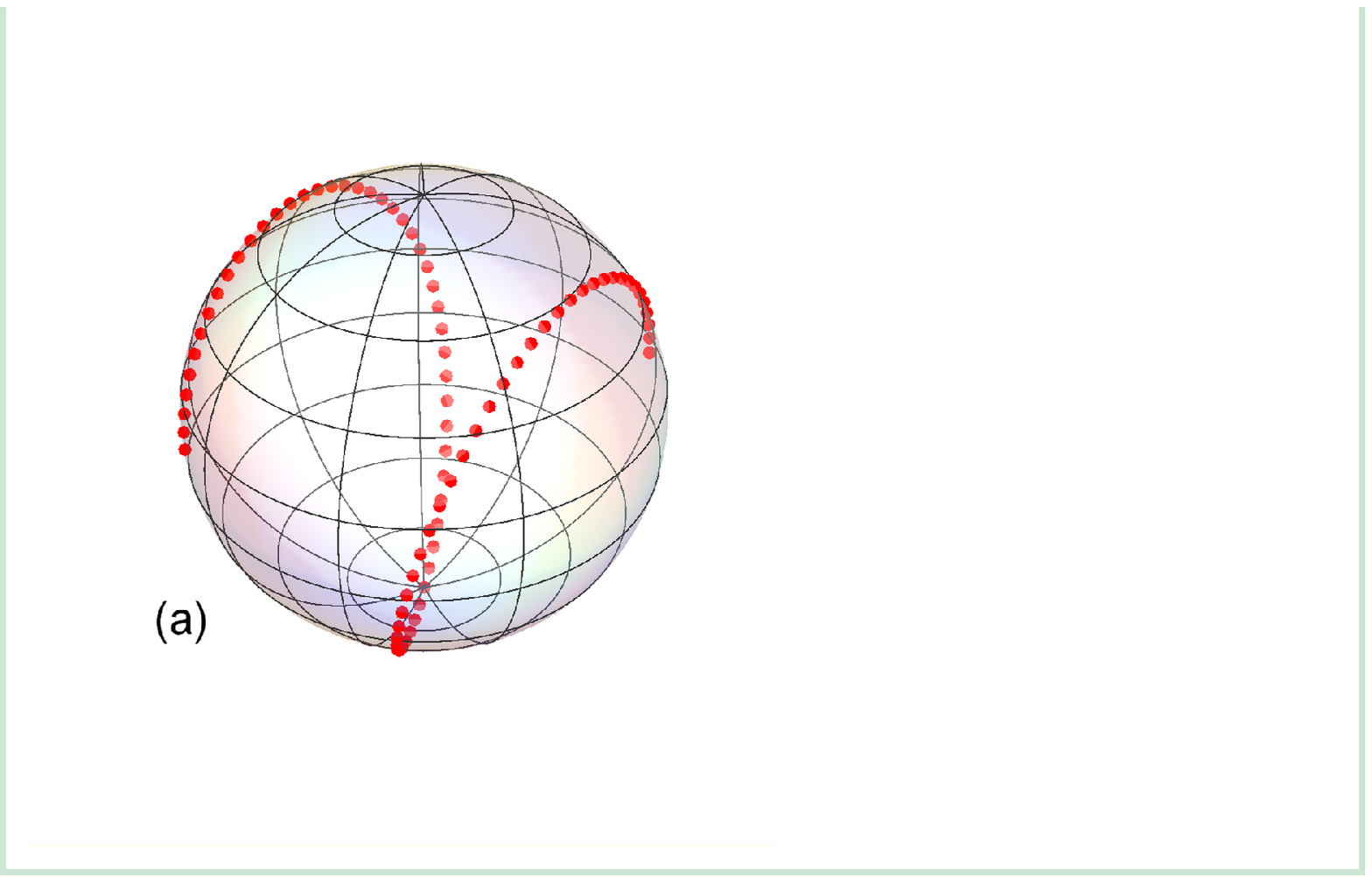}
\includegraphics[width=0.49\columnwidth,trim={40 60 160 40},clip]{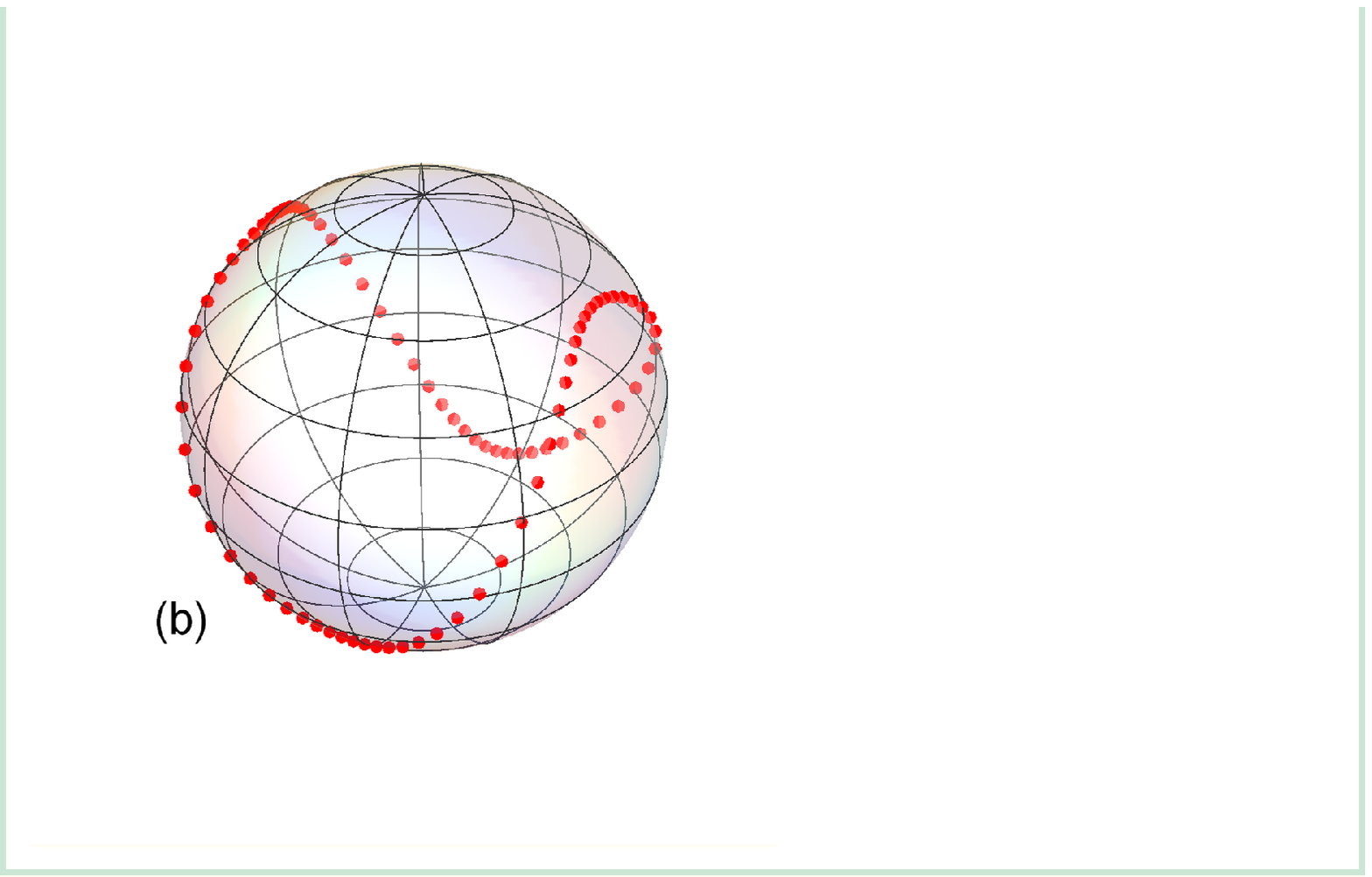}
\end{center}
\caption{(a) The configuration (\ref{e30}) for $d=4$ with $N=80$ nodes, 
omitting
the second component, normalized to a unit 3-vector, showing
equally spaced nodes on $\s^2$, with diametrically opposite endpoints;
(b) the configuration (\ref{f46}) for $d=5$ again with $N=80$ nodes,
with the last component omitted, showing a closed sequence of nodes.
}
\label{fig4}
\end{figure}

For $\kappa_2=0$ the
system synchronizes to a static configuration in which all
nodes lie in a hyperplane defined by 
$\bn\centerdot\bx_i=r_{\infty}=\frac{1}{\sqrt{5}}$
where $\bn=\av/r_{\infty}$, as occurs for $d=3$. These configurations satisfy
\beq
\label{e44}
\bx_i\centerdot\bx_j
=
\frac{1}{5}+\frac{2}{5}\cos\frac{2\pi(i-j)}{N}
+\frac{2}{5}\cos\frac{4\pi(i-j)}{N},
\eeq
from which we deduce that:
\beq
\label{e45}
\bx_i=\frac{1}{\sqrt{5}}
\left(
\sqrt{2}\cos\frac{2\pi i}{N},\sqrt{2}\sin\frac{2\pi i}{N},
\sqrt{2}\cos\frac{4\pi i}{N},\sqrt{2}\sin\frac{4\pi i}{N},1
\right),
\eeq
for which $\av=(0,0,0,0,\frac{1}{\sqrt{5}})$. The equally 
spaced nodes form a closed sequence since $\bx_0=\bx_N$.
We can write these vectors as $\bx_i=\frac{1}{\sqrt{5}}(2\bu_i,1)$
where $\bu_i\in\s^3$ satisfies
$\sum_i\bu_i=0$, corresponding to balanced spacing. Specifically:
\beq
\label{f46}
\bu_i=
\frac{1}{\sqrt{2}}
\left(
\cos\frac{2\pi i}{N},\sin\frac{2\pi i}{N},
\cos\frac{4\pi i}{N},\sin\frac{4\pi i}{N}
\right).
\eeq
These unit vectors are similar to those encountered for $d=4$ in 
(\ref{e30}), which 
lie on a torus in $\s^3$, except that here the nodes form a closed sequence.
We can visualize $\bu_i$ by deleting one component, and normalizing
the remaining components to unity, and then plot the resulting
configuration on $\s^2$, as for $d=4$. As an example, we have plotted
$\bu_i$ in Fig.\ \ref{fig4}(b) for $N=80$, in which the last
component has been dropped, showing a closed sequence of points in $\s^2$, 
in contrast to the $d=4$ case in Fig.\ \ref{fig4}(a).
Under the Hopf fibration (\ref{f37}) we have
$\bu_i\to\left(\cos\frac{6 \pi i}{N},\sin\frac{6\pi i}{N},0\right)$ and so
all nodes are mapped to points on the equator.

\subsection{The 5-body system with 2-body forces}

Of particular interest is the combined system (\ref{e43})
with $\kappa_2\ne0$, by means of which we can investigate
the relative effect of 5-body and $2$-body forces. We find, similar to the
$d=3,4$ cases, that 5-body forces prevail over the $2$-body forces
unless $\kappa_2$
is positive and sufficiently large compared to $|\kappa_5|$, in which
case the system completely synchronizes. The transition to 
complete synchronization is discontinuous, as for $d=4$, except that here we
are able to determine the precise conditions under which this occurs.

The synchronized configuration takes the form
\beq
\label{e47}
\bx_i
=r_{\infty}
\left(
\frac{\alpha}{\sqrt{2}}\cos\frac{2\pi i}{N},
\frac{\alpha}{\sqrt{2}}\sin\frac{2\pi i}{N},
\frac{\alpha}{\sqrt{2}}\cos\frac{4\pi i}{N},
\frac{\alpha}{\sqrt{2}}\sin\frac{4\pi i}{N},1
\right),
\eeq
where the parameters $r_{\infty},\alpha$  satisfy 
$r_{\infty}^2(1+\alpha^2)=1$. This expression
reduces to (\ref{e45}) for $r_{\infty}=\frac{1}{\sqrt{5}}$
and $\alpha=2$. It follows that
\beq
\label{f48}
\bx_i\centerdot\bx_j
=
r_{\infty}^2\left[1+\frac{\alpha^2}{2}\cos\frac{2(i-j)\pi}{N}
+\frac{\alpha^2}{2}\cos\frac{4(i-j)\pi}{N}\right],
\eeq
which shows that the nodes are equally spaced, and since
$\bx_0=\bx_N$ the nodes form a closed sequence.  We require the 
configuration (\ref{e47}) to satisfy the static equations (\ref{g14}), namely:
\beq
\label{e49}
\sum_{j,k,l,m=1}^N\sgn_{ijklm}\,\bv_{jklm}
=
\lambda_1\bx_i-\lambda_2\av,
\eeq
which determines $r_{\infty}$, and hence $\alpha$, as functions
of $\kappa_2/\kappa_5$. We can verify that this equation is indeed 
satisfied, and calculate $\lambda_1,\lambda_2$, see \ref{D}, to obtain
\bea
\label{f50}
\lambda_1
&=&
r_{\infty}(1-r_{\infty}^2)\;
\frac{3 N^2 \cos\frac{2\pi}{N}}{\sin^2\frac{\pi}{N}},
\\
\lambda_2
&=&
 \frac{(1-r_{\infty}^2)\,(5r_{\infty}^2-1)}{r_{\infty}}\;
\frac{3 N^2\cos\frac{2\pi}{N}}{4\sin^2\frac{\pi}{N}}.
\eea
According to (\ref{f15}), the static equations (\ref{e43}) are
satisfied by (\ref{e49}) provided that
$\lambda_2= \kappa_2 N^{4}/\kappa_5$, hence we obtain the following
equation which fixes $r_{\infty}$ as a function of $\kappa_2/\kappa_5$:
\beq
\label{f52}
 \frac{(1-r_{\infty}^2)\,(5r_{\infty}^2-1)}{r_{\infty}}\;
\frac{3\cos\frac{2\pi}{N}}{4N^2\sin^2\frac{\pi}{N}}
=
\frac{\kappa_2}{|\kappa_5|},
\eeq
where we have included the absolute value $|\kappa_5|$ so as to allow for
negative $\kappa_5$, noting that 
the signs of $\kappa_5$ and $\lambda_1,\lambda_2$, change
under $\bx_i\to-\bx_i$.

The configuration (\ref{e47}) exists therefore only if (\ref{f52})
holds. The case $\kappa_2=0$ corresponds to
$r_{\infty}=\frac{1}{\sqrt{5}}$, whereas for $r_{\infty}=1$ we have
$\alpha=\lambda_1=\lambda_2=0$ and (\ref{e47}) becomes an unstable
fixed point. The relation  (\ref{f52}) can be satisfied for
any negative value of $\kappa_2$, however large, corresponding
to arbitrarily small values for $r_{\infty}$, 
and numerically we find indeed that the system synchronizes
such that (\ref{f48}) and (\ref{f52}) are satisfied in all cases.

For positive $\kappa_2$, however, (\ref{f52})
cannot be satisfied if $\kappa_2/|\kappa_5|$ is too large.
Let $f(x)=(1-x^2)(5x^2-1)/x$, then the maximum of $f$ in $(0,1)$
is located at $x^2=(3+2\sqrt{6})/15$, at which $f$ equals
$8(\sqrt{4\sqrt{6}-9})/(3\sqrt{5})\approx1.065$.
This determines the maximum value that $\kappa_2/|\kappa_5|$ can take,
if (\ref{f52}) is to be satisfied.
For any larger values, (\ref{e47}) does not exist as a steady state solution, 
in which case the system
completely synchronizes and so is controlled by the $2$-body forces. 
Evidently, this is a discontinuous
transition, since $r_{\infty}$ jumps from the value
$\approx  0.725$ to unity as $\kappa_2/|\kappa_5|$ increases
through the critical ratio, in contrast to the $d=3$ case.

We conclude that $5$-body forces enhance synchronization
in a way similar to $d=3,4$, in that synchronization occurs
for all coupling constants $\kappa_2,\kappa_5$ in any sign combination.


\section{The symmetric-antisymmetric Kuramoto models\label{s3}}

The simplest of the hierarchy of models (\ref{e6}) is the case $d=2$ which
can be termed the antisymmetric Kuramoto model which, it appears, has
previously escaped attention, but is useful as a guide to 
possible behaviours for $d>2$. We wish to compare the properties for
antisymmetric couplings $\sgn_{ij}$ to those
of the standard Kuramoto model with symmetric couplings $a_{ij}=1$, 
even though in both cases we have $2$-body interactions.
We parametrize $\bx_i=(\cos\theta_i,\sin\theta_i)$, then as defined
in (\ref{f7}), $\bv_i=(\sin\theta_i,-\cos\theta_i)$, and the
potential (\ref{e4}) is given by
$\V_2=\sum_{i,j=1}^N\sgn_{ij}\sin(\theta_j-\theta_i)$.  The equations
of motion are
\beq
\label{e7}
\dot{\theta_i}=\omega_i+\frac{\kappa_a}{N}\sum_{j=1}^N \sgn_{ij}
\cos(\theta_j-\theta_i),
\eeq
where we have included distributed frequencies $\omega_i$, and have
denoted the corresponding coupling constant by $\kappa_a$ which, by
comparison with 
the general system (\ref{e6}), is equal to $-\kappa_d$ for $d=2$.
By writing $\cos(\theta_j-\theta_i)=\sin(\theta_j-\theta_i+\frac{\pi}{2})$
we see that this model is similar to the well-known 
Sakaguchi-Kuramoto model \cite{SK1986} with a frustration angle 
$\alpha=\frac{\pi}{2}$, except that here the couplings are antisymmetric. 
The condition
$|\alpha|<\frac{\pi}{2}$ is imposed in \cite{SK1986} in order to ensure that
the system synchronizes, however the antisymmetric couplings in
(\ref{e7}) allow the system to synchronize for both positive
and negative values for $\kappa_a$, for any frequencies $\omega_i$ and for
any frustration angles, provided that $|\kappa_a|$ exceeds a critical value 
$\kappa_c$. The transformation 
$\theta_i\to-\theta_i$ together with $\omega_i\to-\omega_i$ in
(\ref{e7}) is equivalent to $\kappa_a\to-\kappa_a$, hence the behaviour of the 
system (\ref{e7}) is indifferent  to the sign of $\kappa_a$.  We will see that
the combination
of symmetric and antisymmetric couplings enhances the synchronizability
of the system.


\subsection{Synchronization for identical frequencies\label{ss31}}

\begin{figure}[!ht]
\begin{center}
\includegraphics[width=0.31\columnwidth,trim={10 10 240 10},clip]{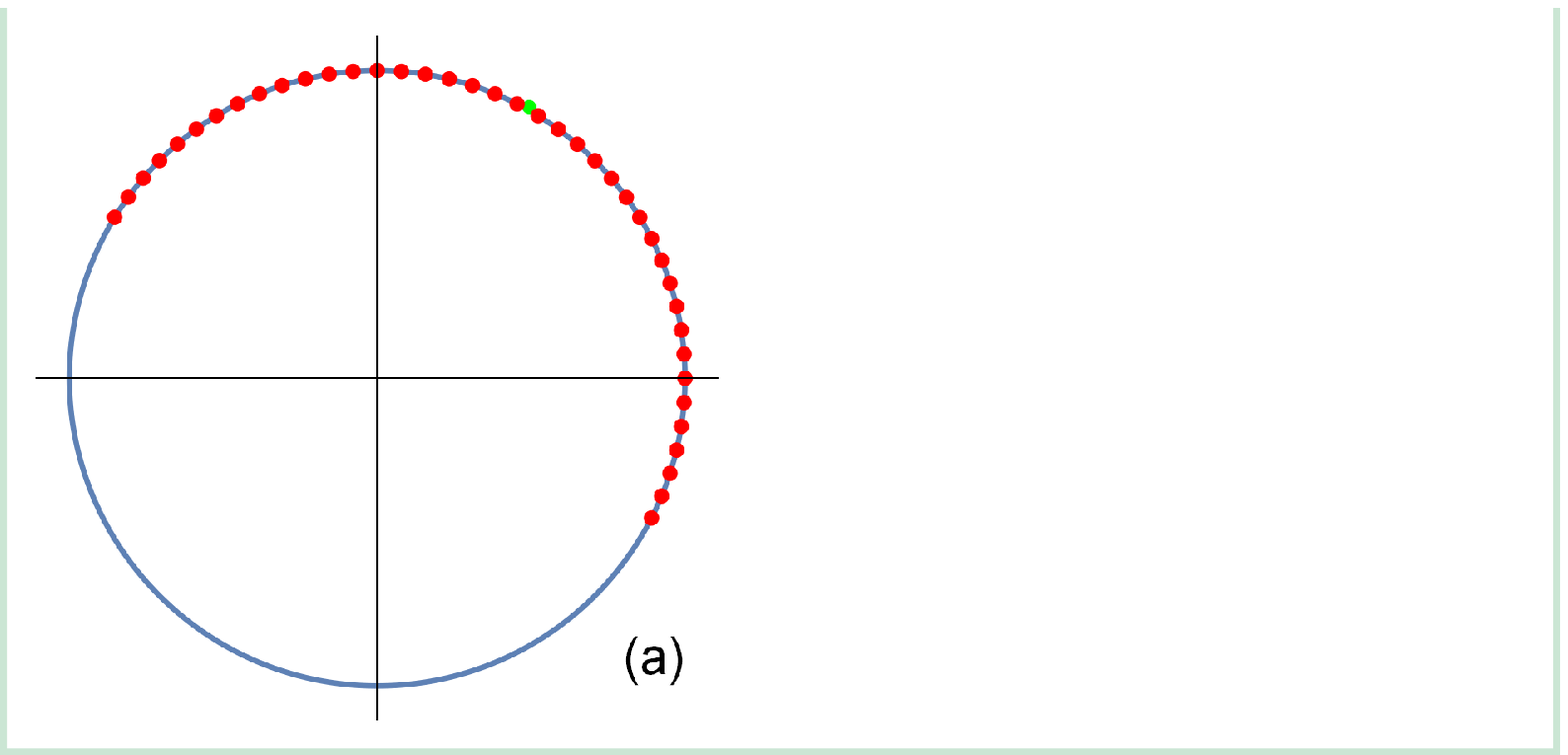}
\includegraphics[width=0.31\columnwidth,trim={10 10 240 10},clip]{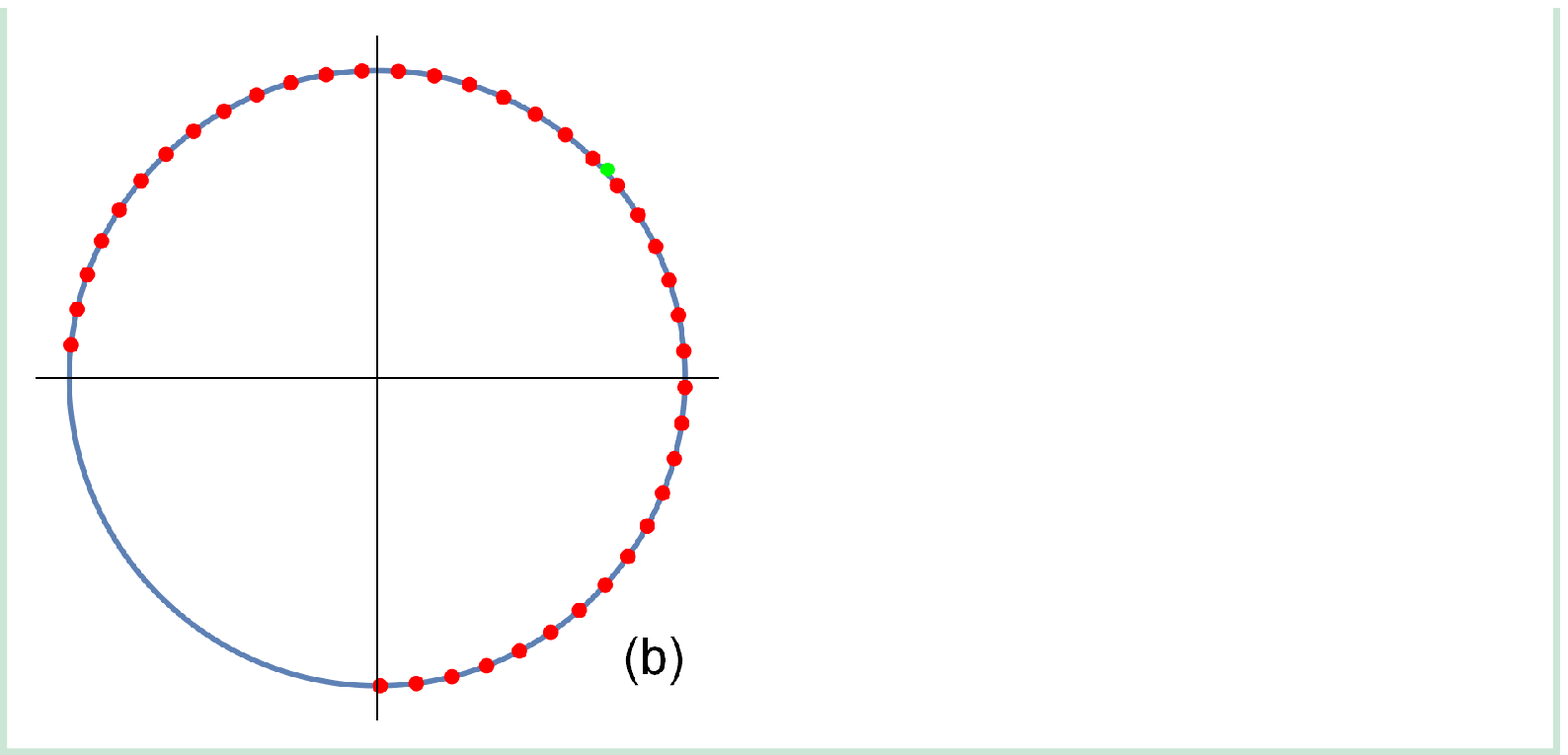}
\includegraphics[width=0.31\columnwidth,trim={10 10 240 10},clip]{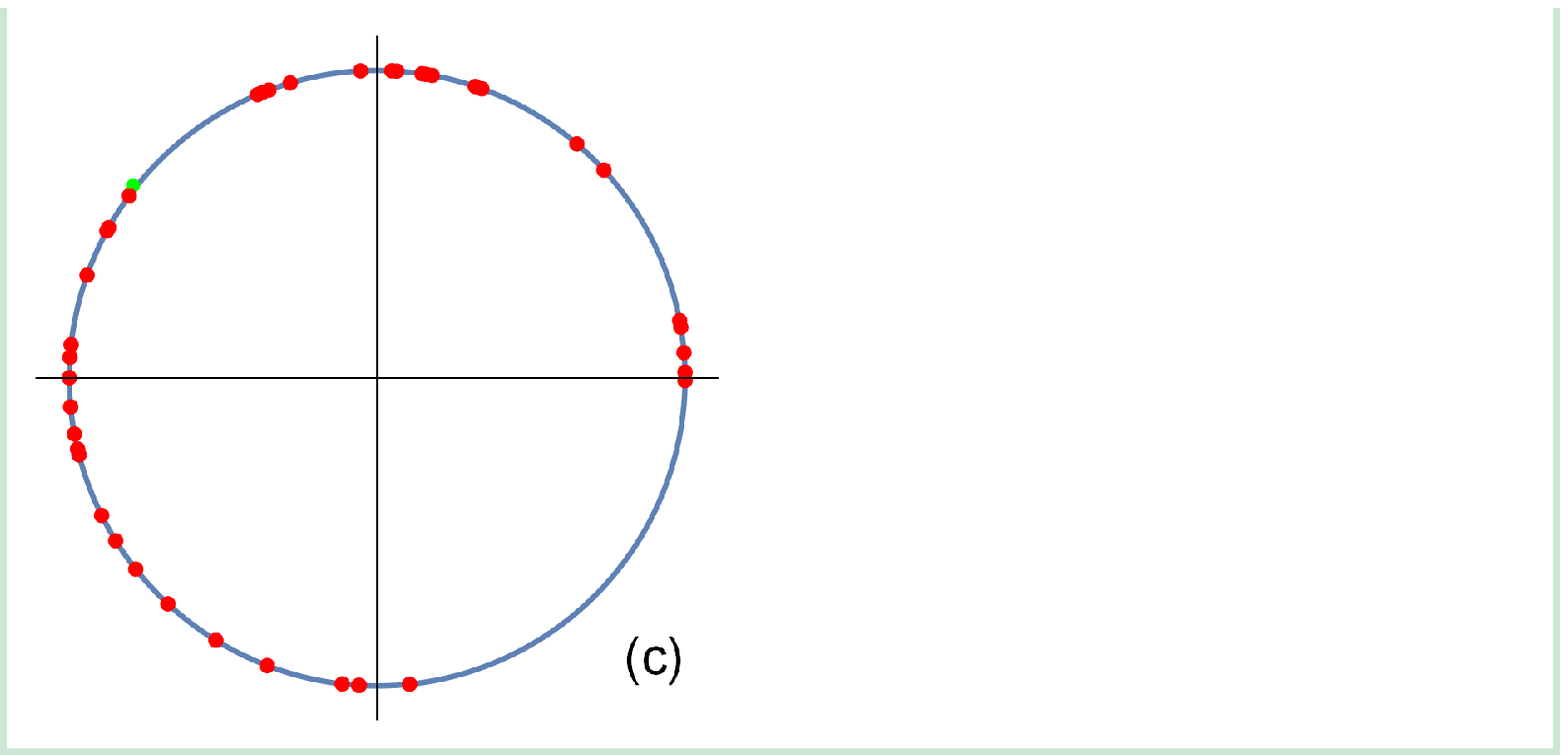}
\end{center}
\caption{
(a) the antisymmetric system (\ref{e7}) with $\omega_i=0$ for $N=40$,
showing the final static configuration with equally spaced nodes on the 
half-circle; (b) the combined system (\ref{e10}) with 
$\omega_i=0,N=40, \kappa_a=-1,\kappa_s=-1$ showing how 
the repulsive interactions for negative $\kappa_s$ spread out the nodes;
(c)  synchronization of (\ref{e10}) for 
$N=40,\kappa_a=\kappa_s=-1$ with distributed nonzero frequencies
$\omega_i$, showing unequally spaced nodes.
}
\label{fig5}
\end{figure}

Taking firstly the case $\omega_i=0$ for all $i$, then we find numerically
that the system (\ref{e7}) synchronizes from random 
initial values $\theta_i(0)$ for any positive or negative $\kappa_a$ 
to a static configuration
in which the nodes are equally spaced on the half circle, as shown for 
example in Fig.\ \ref{fig5}(a) for $N=40$. Such configurations are 
similar to splay states
\cite{Dorfler2014,BS2021}, except that here they are restricted to the 
half-circle.

A steady state solution of (\ref{e7}) 
with $\omega_i=0$ is given by
\beq
\label{e46}
\theta_i=\theta_0+\frac{i\pi}{N}, \qquad i=1,\dots N,
\eeq
where $\theta_0$ is a constant angle, as we now show.
Let $\zeta=\exp\left(\frac{\rmi\pi}{N}\right)$, then 
by means of the geometric series we obtain
\beq
\label{e48}
\sum_{j=1}^N \sgn_{ij}\zeta^{j-i}
=
-\sum_{j=1}^{i-1}\zeta^{j-i}
+\sum_{j=i+1}^N \zeta^{j-i}
=
\frac{1+\zeta}{1-\zeta}
=
\rmi\cot\frac{\pi}{2N},
\eeq
hence 
$\sum_{j=1}^N \sgn_{ij}\cos\frac{(j-i)\pi}{N}=0$ and so (\ref{e46})
is a steady state solution of (\ref{e7}) with $\omega_i=0$.
Let us also verify that the relation (\ref{e12}) holds. We have, choosing
$\theta_0=0$ in (\ref{e46}):
\[
\bx_i=\left(\cos\frac{i\pi}{N},\sin\frac{i\pi}{N}\right), \qquad
\bv_i=\left(\sin\frac{i\pi}{N},-\cos\frac{i\pi}{N}\right),
\]
then from (\ref{e48})
we deduce that $\sum_{j=1}^N\sgn_{ij}\bv_j=\cot\frac{\pi}{2N}\,\bx_i$, 
in accordance with (\ref{e12}).
The value of $r_{\infty}$ is given by
\beq
\label{e9}
r_{\infty}=\frac{1}{N}\Big|\sum_{j=1}^N\rme^{\rmi\theta_j}\Big|
=
\frac{1}{N}\Big|\sum_{j=1}^N\zeta^j\Big|
=
\frac{1}{N}\Big|\frac{2\zeta}{1-\zeta}\Big|
=\frac{1}{N\sin\frac{\pi}{2N}}.
\eeq
Numerical results show that the steady state (\ref{e46}), 
or its negative depending on the sign of $\kappa_a$, is stable 
since it is attained from generic (random) initial values.  The
fixed point $\theta_i=\theta^0$ corresponds
to a completely synchronized configuration but is evidently unstable,
 consistent with  previous observations for $d>2$.
There is some similarity with the case $d=4$ which we considered
in Section \ref{s5}, since in both cases $r_{\infty}$ depends
on $N$, see (\ref{g37}), and the synchronized nodes form a closed loop only with 
the anti-periodic extension in which we include the nodes $-\bx_i$.

For distributed nonzero frequencies $\omega_i$ the system again 
synchronizes provided
that $|\kappa_a|>\kappa_c$ for some critical value $\kappa_c$, where $\kappa_a$
can be positive or negative, with a phase-locked frequency
$\Omega=\frac{1}{N}\sum_j\omega_j$. 
The particles in such a synchronized configuration are no longer exactly
spaced apart but, as $\kappa$ increases, the configuration
approaches the steady state shown in (\ref{e46}).


\subsection{Combined symmetric-antisymmetric models\label{ss72}}

Of particular interest is the effect of the antisymmetric
couplings in  (\ref{e7}) compared to the standard symmetric interactions in 
the  Kuramoto model, and whether the antisymmetric couplings
enhance synchronization, as occurs for 3-body interactions.
Indeed, we find that the antisymmetric interactions dominate the usual symmetric
interactions, for example whereas the Kuramoto model does not synchronize for
negative coupling constants, this is over-ridden by antisymmetric
couplings of any sign, for which the system always synchronizes
to a final configuration with the nodes distributed around the circle.
We consider therefore the system
\beq
\label{e10}
\dot{\theta_i}
=
\omega_i
+\frac{\kappa_s}{N}\sum_{j=1}^N \sin(\theta_j-\theta_i)
+\frac{\kappa_a}{N}\sum_{j=1}^N \sgn_{ij}\cos(\theta_j-\theta_i),
\eeq
with independent coupling constants $\kappa_s,\kappa_a$ of any sign,
which govern
the relative strengths of the symmetric and antisymmetric interactions
respectively.

Consider firstly the case $\omega_i=0$ for all $i$ then, as is well-known,
if $\kappa_a=0$ the system completely synchronizes for $\kappa_s>0$, i.e.\
all particles are co-located at a single point with $r_{\infty}=1$, 
and if $\kappa_s<0$ the 
particles are distributed around the circle such that $r_{\infty}=0$.
However for general $\kappa_s,\kappa_a$, whether positive or negative in any 
combination, the system always synchronizes with the 
final state resembling
that for $\kappa_s=0$, i.e.\ the nodes are equally spaced over an arc of the
unit circle.  The steady state solution in this case is of the form
\beq
\label{e50}
\theta_i=\theta_0 +\frac{i\alpha\pi}{N}, \qquad i=1,\dots N,
\eeq
where $\theta_0$ is a constant angle and $\alpha$ is an unknown
parameter which is determined by
substituting  $\theta_i$ into (\ref{e10}). Let us verify in this case
that the relation (\ref{g14}) is satisfied, and 
hence find the condition which fixes $\alpha$. We have firstly, with the help
of (\ref{aa2},\ref{aa3}):
\beq
\label{e51}
\av=
\frac{1}{N}\sin^2\frac{\alpha\pi}{2}
\left(\cot\frac{\alpha\pi}{2}\cot\frac{\alpha\pi}{2N}-1,\cot\frac{\alpha\pi}{2}
+\cot\frac{\alpha\pi}{2N}
\right),
\eeq
then (\ref{g14}) reads
$\sum_{j=1}^N \sgn_{ij}\bv_{j}
=
\lambda_1\bx_i-\lambda_2\av,$
which is satisfied for all $i$ with
\[
\lambda_1=\cot\frac{\alpha\pi}{2N},
\qquad
\lambda_2=N\cot\frac{\alpha\pi}{2}.
\]
Hence (\ref{e50}) is a steady state solution of (\ref{e10}) provided 
that (\ref{f15}) holds, namely:
\beq
\label{e52}
\frac{\kappa_a}{\kappa_s}
=
-\tan\frac{\alpha\pi}{2},
\eeq
where we have replaced the symmetric coupling coefficient 
$\kappa_2\to\kappa_s$ in (\ref{f15}), together with 
$\kappa_d\to-\kappa_a$, consistent with the sign of 
$\kappa_a$ in (\ref{e10}).
The value of $r_{\infty}$, calculated from (\ref{e51}) or with the help
of (\ref{a4}), is:
\beq
\label{e14}
r_{\infty}
=
\frac{|\sin\frac{\alpha\pi}{2}|}{N|\sin\frac{\alpha\pi}{2N}|},
\eeq
which generalizes the formula (\ref{e9}) for which $\alpha=\pm1$.

The relation (\ref{e52}) shows how the system behaves as the 
coupling ratio $\kappa_s/\kappa_a$ varies, in particular how the spacing 
between nodes changes, as measured by $\alpha$. If $\kappa_s>0$ we set
$\alpha=-\frac{2}{\pi}\arctan\frac{\kappa_a}{\kappa_s}$,
but if $\kappa_s<0$ numerical results show that we must choose either
$\alpha=-$2$-\frac{2}{\pi}\arctan\frac{\kappa_a}{\kappa_s}$ for $\kappa_a>0$,
or $\alpha=$2$-\frac{2}{\pi}\arctan\frac{\kappa_a}{\kappa_s}$ for
$\kappa_a<0$.
The case $\kappa_s=0$ corresponds to $\alpha=\mp1$ 
according to the sign of $\kappa_a$, since either (\ref{e46}) or its 
negative is a stable fixed point.

If $\kappa_s$ is large and positive, for either
positive or negative $\kappa_a$, (\ref{e52}) shows that $\alpha$ 
is small, with $r_{\infty}$ close to unity since from (\ref{e14})
we have $r_{\infty}\to1$ as $\alpha\to0$. Hence the particles are tightly
bunched together, but are always equally spaced. 
Complete synchronization never occurs, even for large values of 
$\kappa_s$, in contrast to the $d=3$ case, see Section \ref{ss51}.
If $\kappa_s$ is small,
whether positive or negative, the configuration is similar to that
when $\kappa_s=0$, since $|\alpha|$ is close to unity. 
If $\kappa_s$ is large and negative, $|\alpha|$ increases, but with
$|\alpha|<2$, and so the nodes spread out to occupy almost the full circle.
As an example, we have plotted such a configuration for $N=40$ in 
Fig.\ \ref{fig5}(b) with $\kappa_a=\kappa_s=-1$, for which
$\alpha=\frac{3}{2}$.
The $\kappa_s$ coupling therefore either increases the length of the arc 
on which the synchronized points lie, corresponding to a repulsive effect 
for $\kappa_s<0$, or binds the nodes together more tightly
for $\kappa_s>0$.

Consider finally the case of distributed frequencies $\omega_i$, then
synchronization occurs with a phase-locked frequency 
$\Omega=\frac{1}{N}\sum_j\omega_j$, but only if at least one of 
$\kappa_a,\kappa_s$ is sufficiently large in magnitude.
The Kuramoto model does not synchronize if the coupling 
coefficient is negative, but now we find that synchronization does indeed occur
if $|\kappa_a|$ is sufficiently large, i.e.\ the antisymmetric
couplings override the repulsion of nodes arising from the negative
symmetric couplings. These synchronized configurations are similar to those
encountered in models with distributed phase lag, which can be
understood if we write (\ref{e10}) in the form
\beq
\label{e15}
\dot{\theta_i}
=
\omega_i
+
\frac{\kappa_s}{N}
\sqrt{1+\frac{\kappa_a^2}{\kappa_s^2}}\;\sum_{j=1}^N 
\sin\left(\theta_j-\theta_i+\sgn_{ij}\,\arctan\frac{\kappa_a}{\kappa_s}
\right).
\eeq
The main difference between this system and the Sakaguchi-Kuramoto
model with distributed phase-lag is that the angles are 
antisymmetric, a consequence of which is that (\ref{e15}) 
has a gradient formulation, unlike the 
Sakaguchi-Kuramoto model for which symmetric angles are usually
assumed.

As an example of a synchronized configuration for the system (\ref{e15}),
equivalently (\ref{e10}), we show in Fig.\ \ref{fig5}(c) a final
synchronized
state with $N=40$ for $\kappa_a=\kappa_s=-5$, with distributed
frequencies such that $|\omega_i|<1$. Since $\kappa_s<0$ this system does
not synchronize unless we include a nonzero $\kappa_a$ coupling of
sufficiently large magnitude, and then synchronization occurs even 
if $\kappa_s$ is very large and negative.


\section{Conclusion\label{s8}}

We have constructed a hiercharchy of models on the unit sphere 
in $d$ dimensions which have either exclusive $d$-body interactions or combined 
$d$- and $2$-body interactions. We have shown that such systems 
synchronize under very general conditions to a steady state
with equally spaced particles on $\s^{d-1}$, 
unless the $2$-body forces are sufficiently
strong in which case the system completely synchronizes.
We have derived exact expressions for the synchronized configurations 
for $d\leqslant5$, and have calculated for $d=3,5$ 
the exact ratio of coupling constants
at which the system transitions to a completely synchronized
configuration. Our findings
support previous observations that higher-order interactions enhance
the synchronization of the combined systems even when the pairwise couplings
are repulsive, but we have not observed other reported phenomena such as
multistability or explosive synchronization.

Some of the properties that we have derived for $d\leqslant5$
generalize without difficulty to any $d$, for example in the simplest case 
in which $N=d$, an exact static solution of the $d$-body system (\ref{e6})
is $\bx_i=\be_i$, where 
$\{\be_1,\dots \be_d\}$ is the standard basis in $\R^d$, in which the vectors
$\bx_i$ are aligned along the orthogonal axes, and hence
$r_{\infty}=1/\sqrt{d}$, consistent with $d=3,5$, also $d=4$, see
(\ref{g37}) with $N=4$, and $d=2$, see (\ref{e9}) with $N=2$. 
One would also expect that the exact expressions for steady state 
solutions, such as (\ref{e16}) for $d=3$, (\ref{e30}) for $d=4$,
and (\ref{e45}) for $d=5$, extend to larger values of $d$ for general $N$.  

Our aim has been to determine the dynamical characteristics
of higher-order interactions with antisymmetric coupling
coefficients, and it remains to establish properties of more general
couplings with their associated network structures.  There are also 
many mathematical questions to be finalized, such as the stability
of the asymptotic configurations, as well as possible applications in this vast
area of research.


\appendix
\section{\label{A}}
\setcounter{section}{1}

We derive here several well-known trigonometric summation formulas.
Let $\zeta=\exp\left(\frac{\rmi\pi}{N}\right)$ then $\zeta^N=-1$, and
from the geometric series,
$\sum_{k=p}^q x^k=(x^p-x^{q+1})/(1-x)$,
for any parameter $\alpha$:
\beq
\label{a1}
\sum_{j=1}^N \rme^{\rmi \alpha\pi j/N}
=
\sum_{j=1}^N (\zeta^{\alpha})^j
=
\frac{\zeta^{\alpha}(1-\rme^{\rmi\alpha\pi})}{1-\zeta^{\alpha}},
\eeq
and so
\bea
\label{aa2}
\sum_{j=1}^N\cos\frac{\alpha\pi j}{N}
&=&
\half\left(-1+\cos\alpha\pi+\cot\frac{\alpha\pi}{2N}\sin\alpha\pi  \right),
\\
\label{aa3}
\sum_{j=1}^N\sin\frac{\alpha\pi j}{N}
&=&
\frac{\sin\frac{\alpha\pi}{2}\sin\frac{\alpha(N+1)\pi}{2N}}
{\sin\frac{\alpha\pi}{2N}}.
\eea
For even integers $m$, therefore
\beq
\label{a2}
\sum_{j=1}^N\cos\frac{m\pi j}{N}
=0
=
\sum_{j=1}^N\sin\frac{m\pi j}{N},
\eeq
and for odd integers $m$:
\beq
\label{a3}
\sum_{j=1}^N \cos\frac{m\pi j}{N}=-1, \qquad
\sum_{j=1}^N \sin\frac{m\pi j}{N}=\cot\frac{m\pi}{2N}.
\eeq
From (\ref{a1}) we also obtain:
\beq
\label{a4}
\fl
\sum_{i,j=1}^N \rme^{\rmi \alpha\pi (j-i)/N}
=
\sum_{i=1}^N(\zeta^{\alpha})^{-i} \sum_{j=1}^N(\zeta^{\alpha})^{j}
=
\frac{(1-\rme^{-\rmi\alpha\pi})(1-\rme^{\rmi\alpha\pi})}
{(1-\zeta^{-\alpha})(1-\zeta^{\alpha})}
=
\frac{\sin^2\frac{\pi\alpha}{2}}{\sin^2\frac{\pi\alpha}{2N}},
\eeq
which, since the right-hand side is real, is equal to
$\sum_{i,j=1}^N \cos\frac{\alpha\pi (j-i)}{N}$.


\section{\label{B}}
\setcounter{section}{2}

Here we prove that (\ref{e17}) and (\ref{e23}) are satisfied by the 
vectors (\ref{e16}). Firstly, it follows from (\ref{e16}) that
\beq
\label{b1}
\fl
\bx_j\times\bx_k
=
\frac{\sqrt{2}}{3}\left(
\sin\frac{2j\pi}{N}-
\sin\frac{2 k\pi}{N},
-\cos\frac{2j\pi}{N}+
\cos\frac{2k\pi}{N},
-\sqrt{2}\sin\frac{2(j-k)\pi}{N}\right),
\eeq
and we wish to evaluate $\sum_{j,k=1}^N\sgn_{ijk}(\bx_j\times\bx_k)$.
We first evaluate $\sum_{k=1}^N\sgn_{ijk}\,\zeta^{-2k}$ where
$\zeta=\exp\left(\frac{\rmi\pi}{N}\right)$. If $i<j$ then
$\sgn_{ijk}=1$ for  $1\leqslant k<i$,  $\sgn_{ijk}=-1$ for $i<k<j$
and $\sgn_{ijk}=1$ for $j<k\leqslant N$.
By means of the geometric series we obtain for $i<j$:
\beq
\label{b2}
\fl
\sum_{k=1}^N\sgn_{ijk}\,\zeta^{-2k}
=
\sum_{k=1}^{i-1}\zeta^{-2k}
-\sum_{k=i+1}^{j-1}\zeta^{-2k}
+\sum_{k=j+1}^N\zeta^{-2k}
=
\frac{(1+\zeta^2)(\zeta^{-2i}-\zeta^{-2j})}{1-\zeta^2},
\eeq
which is antisymmetric in $i,j$, and therefore holds for all $i,j$. 
By summing over $j$ and using $\sum_{j=1}^N\zeta^{-2j}=0$,
as follows from (\ref{a1}) with $\alpha=-2$, we obtain:
\[
\frac{1}{N}\sum_{j,k=1}^N\sgn_{ijk}\,\zeta^{-2k}
=
\frac{\zeta^{-2i}(1+\zeta^2)}{1-\zeta^2}
=
\rmi \zeta^{-2i} \cot\frac{\pi}{N}.
\]
Hence:
\[
\frac{1}{N}\sum_{j,k=1}^N\sgn_{ijk}\sin\frac{2k\pi}{N}
=
-\cot\frac{\pi}{N}\cos\frac{2\pi i}{N}
=
-\frac{1}{N}\sum_{j,k=1}^N\sgn_{ijk}\sin\frac{2j\pi}{N},
\]
where we swapped the $j,k$ summations, as well as
\[
\frac{1}{N}\sum_{j,k=1}^N\sgn_{ijk}\cos\frac{2k\pi}{N}
=
\cot\frac{\pi}{N}\sin\frac{2\pi i}{N}
=
-\frac{1}{N}\sum_{j,k=1}^N\sgn_{ijk}\cos\frac{2j\pi}{N}.
\]
Also, from (\ref{b2}):
\[
\frac{1}{N}\sum_{j,k=1}^N\sgn_{ijk}\,\zeta^{2j-2k}
=
\frac{1}{N}\sum_{j=1}^N\zeta^{2j} \sum_{k=1}^N\sgn_{ijk}\,\zeta^{-2k}
=
-\frac{1+\zeta^2}{1-\zeta^2}
=
-\rmi\cot\frac{\pi}{N},
\]
again using $\sum_{j=1}^N\zeta^{2j}=0$, and so
\[
\frac{1}{N}\sum_{j,k=1}^N\sgn_{ijk}\sin\frac{2(j-k)\pi}{N}
=
-\cot\frac{\pi}{N}.
\]
By application of the above formulas we evaluate
$\sum_{j,k=1}^N\sgn_{ijk}(\bx_j\times\bx_k)$ using (\ref{b1}) 
to obtain (\ref{e17}) and (\ref{e23}).


\section{\label{D}}
\setcounter{section}{3}

We prove here (\ref{e35}) for $d=4$ and similar relations
for $d=5$. We first calculate the components
of $\bv_{jkl}$ from the identity
$\bu\centerdot\bv_{jkl}= \det(\bu,\bx_{j},\bx_k, \bx_{l})$
for an arbitrary vector $\bu$ and find for the first component,
with the help of a computer algebra system:
\bea
\nonumber
\fl
4\sqrt{2}\,(\bv_{jkl})^1
&=&
\cos\frac{(3j-k-3l)\pi}{N}
-
\cos\frac{(3j+k-3l)\pi}{N}
+
\cos\frac{(j+3k-3l)\pi}{N}
\\
\label{d1}
&-&
\cos\frac{(3j-3k-l)\pi}{N}
+
\cos\frac{(3j-3k+l)\pi}{N}
-
\cos\frac{(j-3k+3l)\pi}{N},
\eea
and similarly for the other three components. We wish to evaluate
$\sum_{j,k,l=1}^N \sgn_{ijkl}\bv_{jkl}$. For the first component  
we obtain from (\ref{d1}):
\beq
\label{d2}
4\sqrt{2}\,\sum_{j,k,l=1}^N\sgn_{ijkl}\,(\bv_{jkl})^1
=
6\sum_{j,k,l=1}^N\sgn_{ijkl}\,\cos\frac{(3j-k-3l)\pi}{N}
\eeq
where we have used the antisymmetric properties of 
$\sgn_{ijkl}$ to combine the six terms on the right-hand side of (\ref{d1})
into a single term.
In order to evaluate this sum, let
$\zeta=\exp\left(\frac{\rmi\pi}{N}\right)$ then with
methods similar to those in \ref{B} we derive
\beq
\label{d3}
\sum_{k,l=1}^N\sgn_{ijkl}\,\zeta^{\alpha k}\zeta^{\beta l}
=
\frac
{(1+\zeta^{\alpha})(1+\zeta^{\beta})}
{(1-\zeta^{\alpha})(1-\zeta^{\beta})}
\left(\zeta^{\alpha j+\beta i}-\zeta^{\alpha i+\beta j}\right),
\eeq
which is valid for all $i,j$ and all odd integers $\alpha,\beta$ such that
$\alpha+\beta\ne0$. For $\alpha=-1,\beta=-3$ we obtain
\bea
\nonumber
\sum_{j,k,l=1}^N\sgn_{ijkl}\,\zeta^{3j-k-3l}
&=&
\frac
{(1+\zeta^{-1})(1+\zeta^{-3})}
{(1-\zeta^{-1})(1-\zeta^{-3})}
\sum_{j=1}^N\left(\zeta^{2j-3i}
-\zeta^{- i}\right)
\\
\label{d5}
&=&
-N\frac
{(1+\zeta^{-1})(1+\zeta^{-3})\zeta^{- i}}
{(1-\zeta^{-1})(1-\zeta^{-3})},
\eea
where
\[
\frac
{(1+\zeta^{-1})(1+\zeta^{-3})}
{(1-\zeta^{-1})(1-\zeta^{-3})}
=
-\cot\frac{\pi}{2N}\cot\frac{3\pi}{2N}.
\]
The real part of (\ref{d5}) now reads:
\beq
\label{d6}
\sum_{j,k,l=1}^N\sgn_{ijkl}\,
\cos\frac{(3j-3l-k)\pi}{N}
=
N \cot\frac{\pi}{2N}\cot\frac{3\pi}{2N}\cos\frac{\pi i}{N},
\eeq
which establishes the first component of (\ref{e35}). The second
component is verified by taking the imaginary part of
(\ref{d5}), and a similar calculation holds for the two remaining components.

For the more general expression (\ref{e40}) for $\bx_i$, which includes
the effect of $2$-body forces, we calculate the components
of $\bv_{jkl}$ as before, and again find that each component is a sum of 
six terms,
generalizing (\ref{d1}). By means of the antisymmetry properties of 
$\sgn_{ijkl}$ we obtain an expression which reduces to (\ref{d2})
for $\theta=\frac{\pi}{4},\alpha=\beta =1$, namely:
\beq
\label{d7}
\sum_{j,k,l=1}^N\sgn_{ijkl}\,(\bv_{jkl})^1
=
3\cos\theta\sin^2\theta\sum_{j,k,l=1}^N\sgn_{ijkl}\,
\cos\frac{(3\beta j-3\beta l -\alpha k)\pi}{N}.
\eeq
We can evaluate this expression in principle by summing 
$\sum_{j,k,l=1}^N\sgn_{ijkl}\,\zeta^{3\beta j-3\beta l -\alpha k}$, and
then extracting the real part, however
because $\alpha,\beta$ are not integers, the result does not
simplify to an expression such as (\ref{d5}). Due to the resulting
complexity we resort
to numerical observations for this case.

For $d=5$ we wish to prove that (\ref{e47}) satisfies (\ref{e49}), and
find explicit expressions for $\lambda_1,\lambda_2$. We firstly
evaluate $\bv_{jklm}$ by means of
$\bu\centerdot\bv_{jklm}= \det(\bu,\bx_{j},\bx_k, \bx_{l},\bx_m)$ again
with the help of a computer algebra system,
to obtain an expression for each component,
similar to (\ref{d1}) for $d=4$, except that there are now 24 terms, instead
of 6, on the right-hand side.
When summed over the signature, however, these 24 terms can be reduced
to a single term, similar to (\ref{d2}) and (\ref{d7}) for $d=4$, 
by using the antisymmetric properties of the signature function, 
 hence we obtain
\beq
\fl
\label{d8}
\sum_{j,k,l,m=1}^N\sgn_{ijklm}\,(\bv_{jklm})^1
=
3 r_{\infty}^4\alpha^3\sqrt{2}
\sum_{j,k,l,m=1}^N\sgn_{ijklm}\,\cos\frac{(4j-2k-4l)\pi}{N}.
\eeq
It is sufficient to consider only the first component of 
(\ref{e49}) in order to find $\lambda_1$, since we have
$\av=(0,0,0,0,r_{\infty})$. Having found $\lambda_1$ we then determine
$\lambda_2$ by considering the 5th component of (\ref{e49}). It is easiest
to first find $\lambda_1$ by setting $i=0$ in (\ref{d8}), then 
by means of sums such as (\ref{d3}) we obtain:

\[
\sum_{j,k,l,m=1}^N\sgn_{jklm}\,\cos\frac{(4j-2k-4l)\pi}{N}
=
\frac{N^2}{2}\frac{\cos\frac{2\pi}{N}}{\sin^2\frac{\pi}{N}},
\]
from which we can identify $\lambda_1$ as given in (\ref{f50}), and similarly
obtain $\lambda_2$. Expressions such as (\ref{e49}) can also be verified 
exactly for specific, but small, $N$ by means of a computer algebra system,
as well by high-accuracy numerical computations.


\section{\label{C}}
\setcounter{section}{4}

We show here how to rotate static configurations, 
such as that shown in Fig.\ \ref{fig1}(a), to a fixed orientation so as to
directly compare final states for different initial values.
In particular we wish to rotate $\av$ as defined in (\ref{f9}),
equivalently the normal $\bn=\av/\|\av\|$, to the vector 
$\mm=(0,0,\dots,1)$.

The following method performs the required transformation in any dimension $d$ 
by means of a specific $\mathrm{SO}(d)$ rotation, 
after which we still have the freedom
to perform $\mathrm{SO}(d-1)$ rotations about $\mm$.
Let $\bw=(\omega_1,\omega_2,\dots \omega_{d-1})$ be any column
vector of length $d-1$, and define the $d\times d$ 
antisymmetric matrix
\beq
\Omega=
\pmatrix{
0 &\bw
\cr
-\bw^{\tp} &0},
\eeq
where the upper left block is the zero matrix of size  $(d-1)\times (d-1)$, 
and $\bw^{\tp}$ denotes the transpose of $\bw$. Then
\[
\Omega^2=
\pmatrix{
-\bw\bw^{\tp}&0
\cr
0&-\omega^2},
\]
where $\omega=\sqrt{\bw\centerdot\bw}$, and hence $\Omega^3=-\omega^2\,\Omega$.
In order to evaluate the rotation matrix $R=\rme^{-\Omega}$
we expand the matrix exponential according to
$\rme^{t\,\Omega}=a(t)I_d+b(t)\Omega+c(t)\Omega^2$ where
$a(0)=1,b(0)=0,c(0)=0$ then, on differentiating, we obtain
\[
\rme^{t\,\Omega}\Omega=\dot{a}\,I_d+\dot{b}\,\Omega +\dot{c}\,\Omega^2
=
a\Omega+b\Omega^2+c\Omega^3
=
a\Omega+b\Omega^2-c\omega^2\Omega.
\]
This implies that $\dot{a}=0, \dot{b}=a-\omega^2 c,\dot{c}=b$ and hence
we obtain
$a=1, b=\sin\omega t/\omega, c=(1-\cos\omega t)/\omega^2$, which
gives an explicit expression for 
$\rme^{t\,\Omega}= I_d+b(t)\Omega+c(t)\Omega^2$.
In particular:
\[
\rme^{t\Omega}\mm
=
\pmatrix{\frac{\sin\omega t}{\omega}\bw\cr \cos\omega t}.
\]
By equating the right-hand side of this expression at $t=1$
with the computed normal $\bn$, we can identify $\bw$, and therefore obtain
the rotation matrix $R$ with the required property $R\bn=\mm$, where
$R$ is given in explicit form by $R=\rme^{-\Omega}$. Hence 
we can rotate a configuration $\bx_i$ at any fixed time for all $i$ according to 
$\bx_i\to R\bx_i$ with the property that $\av$ aligns with $\mm$.



\end{document}